\documentclass[11pt]{article}
\pdfoutput=1
\usepackage{jheppub}
\usepackage{graphicx}
\usepackage{amssymb}
\usepackage{amsmath,amssymb}
\usepackage{slashed}
\usepackage{hyperref}
\usepackage{caption}
\usepackage{xcolor}
\usepackage{dsfont}
\usepackage{verbatim}
\usepackage{subfig}
\usepackage{amsmath,amssymb,amscd,amsfonts,mathtools}
\usepackage{xcolor}
\definecolor{markgreen}{RGB}{230,243,230}
\definecolor{darkolivegreen}{rgb}{0.33, 0.42, 0.18}
\definecolor{darkpastelgreen}{rgb}{0.01, 0.75, 0.24}
\DeclareMathOperator{\Tr}{Tr}

\usepackage[toc,page]{appendix}
\usepackage{epsfig}
\usepackage{epstopdf}
\usepackage{latexsym}
\usepackage{graphicx}
\usepackage{booktabs}
\usepackage{bbm}
\usepackage{color}
\usepackage{float}
\usepackage{physics}
\usepackage{tensor}
\usepackage{verbatim}
\usepackage{subfig}
\usepackage{tikz}
\usepackage{ifthen}
\usetikzlibrary{matrix}
\usetikzlibrary{decorations.markings,calc,shapes,decorations.pathmorphing,patterns,decorations.pathreplacing,arrows.meta}
\usetikzlibrary{positioning}
\usepackage{hyperref}
\usetikzlibrary{patterns}
\usetikzlibrary{positioning}

\pdfoutput=1
\makeatletter
\def\@fpheader{\relax}
\makeatother

\newboolean{showcomments}
\setboolean{showcomments}{false}
\newcommand\rem[1]{\ifthenelse{\boolean{showcomments}}{{#1}}{}}
\newcommand{\be}{\begin{equation}}
\newcommand{\ee}{\end{equation}}

\title{\Large Microscopic Origin of the Entropy of Single-sided Black Holes}
\author{Hao Geng$^a$, }
\author{Yikun Jiang$^b$}
\affiliation{$^a$Center for the Fundamental Laws of Nature, Harvard University, 17 Oxford St., Cambridge, MA, 02138, USA.}
\affiliation{$^b$Department of Physics, Northeastern University, Boston, MA 02115, USA
}
\emailAdd{haogeng@fas.harvard.edu}
\emailAdd{phys.yk.jiang@gmail.com}
\abstract{In this paper, we provide a state-counting derivation of the Bekenstein-Hawking entropy formula for single-sided black holes. We firstly articulate the concept of the black hole microstates. Then we construct explicit mircostates of single-sided black holes in (2+1)-dimensional spacetimes with a negative cosmological constant. These microstates are constructed by putting a Karch-Randall brane behind the black hole horizon. Their difference is described by different interior excitations which gravitationally backreact. We show that these microstates have nonperturbatively small overlaps with each other. As a result, we use this fact to give a state-counting derivation of the Bekenstein-Hawking entropy formula for single-sided black holes. At the end, we notice that there are no negative norm states in the resulting Hilbert space of the black hole microstates which in turn ensures unitarity.
All calculations in this paper are analytic and can be easily generalized to higher spacetime dimensions.
}
\begin{document}
\maketitle
\begin{sloppypar}
\flushbottom
\pagebreak
\section{Introduction}\label{sec:intro}
Black holes were discovered at the very early days of the general theory of relativity as nontrivial solutions of the vacuum Einstein's field equation. They are characterized by only a few macroscopic parameters. Nevertheless, it was found by Hawking \cite{Hawking:1974sw} that black holes exhibit interesting phenomena at the quantum level. These phenomena already manifest at the tree-level that black holes radiate thermal particles if one turns on any probe quantum fields on the black hole background. This observation further suggests that black holes are thermodynamical systems obeying the basic laws of thermodynamics \cite{Bardeen:1973gs}. The most remarkable implication of the thermodynamical nature of black holes is that a black hole carries an entropy \cite{Bekenstein:1973ur,Hawking:1974sw,Hawking:1976ra} 
\begin{equation}
    S_{\text{BH}}=\frac{A}{4G_{N}}\,,\label{eq:BH}
\end{equation}
where $A$ is the area of the black hole horizon and $G_{N}$ is Newton's constant. This formula is rather universal as it works for any kind of black holes in general relativity and it is called the Bekenstein-Hawking entropy formula. Nevertheless, this formula suggests that black holes are rather exotic systems whose entropy is not proportional to their volume but instead their area. Moreover, the statistical nature of this formula has been a long-standing puzzle which suggests that a proper understanding of the quantum theory of gravity is desired.

\begin{figure}
	\centering	\includegraphics[width=0.7\linewidth]{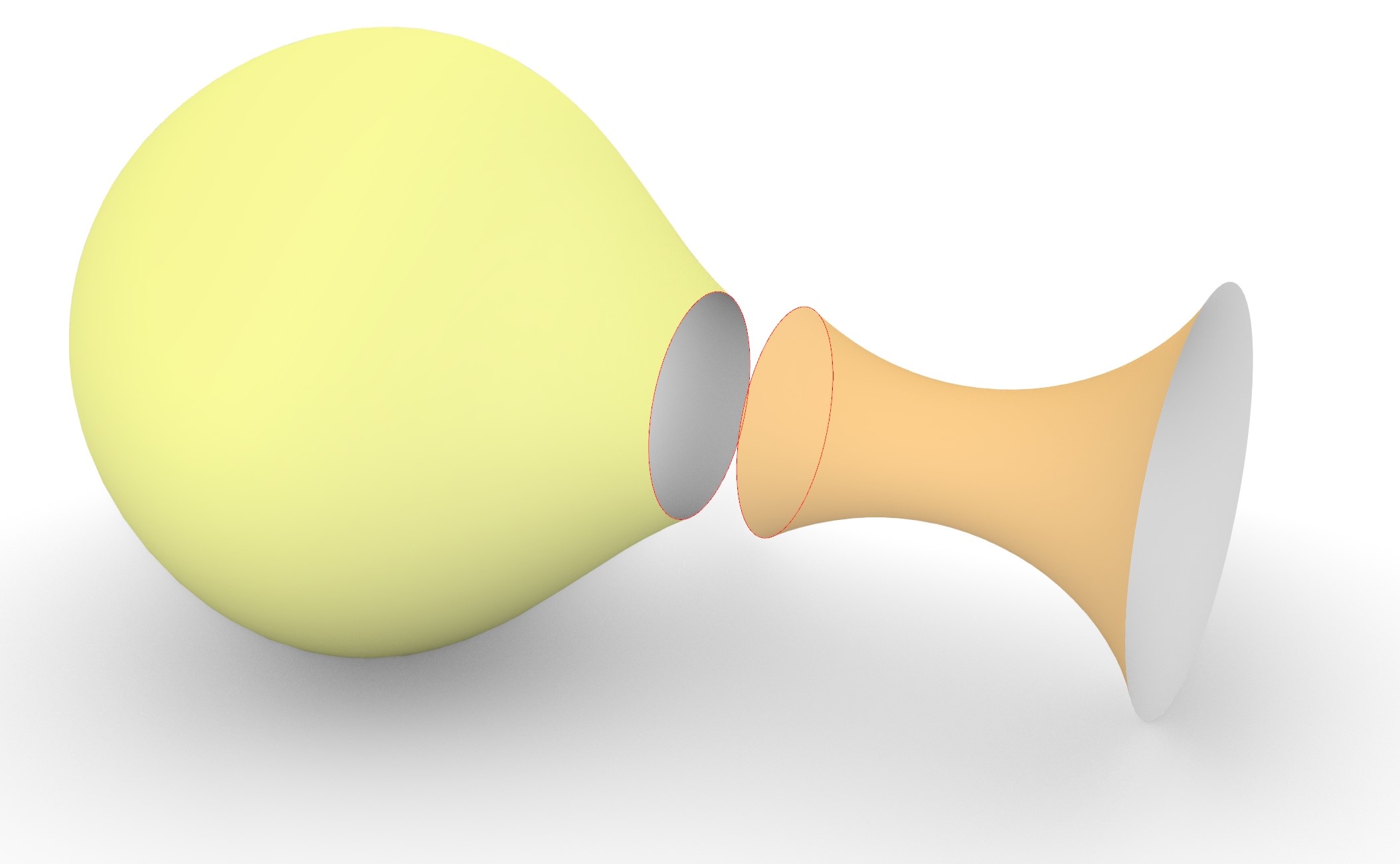}
	\caption{\small A cartoon of how a bag of gold can be constructed. The yellow cap is part of the spatial section of an FRW universe and it is glued to the trumpet along the red circle in a $\mathcal{C}^{1}$ continuous way. The trumpet is the spatial section of a black hole spacetime. The bell of the trumpet is the asymptotic boundary of the black hole spacetime. The neck of the trumpet is the black hole horizon. Therefore, the FRW universe is behind the black hole horizon.}
	\label{pic:bog}
\end{figure}

The quantum statistical mechanics interpretation of the formula Equ.~(\ref{eq:BH}) is that the number of linearly independent black hole microstates, which have identical macroscopic properties as the black hole solution, should be exactly 
\begin{equation}
\Omega_{\text{BH}}=e^{S_{\text{BH}}}=e^{\frac{A}{4G_{N}}}\,,\label{eq:Omega}
\end{equation}
where in the full quantum theory $G_{N}$ should be the renormalized Nezwton's constant.\footnote{In this paper, we don't consider higher derivative corrections though one should notice that these corrections are naturally generated at the quantum level \cite{Solodukhin:2004gs}.} Nevertheless, one can easily construct paradoxes that are naively against the above interpretation. The most famous one among such paradoxes is the bag of gold paradox.\footnote{The earliest papers we are able to find pointing out this paradox against the statistical interpretation of the Bekenstein-Hawking entropy formula are \cite{Hsu:2008yi,Marolf:2008tx}.} The bag of gold is a specific solution of Einstein's field equation that looks the same as a black hole from the exterior of its horizon but is different inside. This solution is originally constructed by Wheeler \cite{wheeler1964relativity} where Wheeler glues an FRW universe behind the horizon of the black hole on the zero-time slice (i.e.~the symmetric slice of time evolution) with the full geometry as the time evolution of this initial condition (see Fig.~\ref{pic:bog}). This solution naively contradicts the above statistical interpretation of the Bekenstein-Hawking entropy formula as the FRW universe can be very large and one can arbitrarily excite the black hole interior by exciting the quantum fields in this FRW universe. This naively gives a large number of states which scale with the volume of the FRW universe, and they look the same as the black hole outside the horizon. Moreover, to prove the statistical interpretation of the Bekenstein-Hawking entropy formula one may want to construct a complete set of microstates and show that there are exactly $e^{\frac{A}{4G_{N}}}$ linearly independent states among them. However, it was unclear how such set of microstates can be constructed and counted without invoking fine details of the gravitational theory \cite{Strominger:1996sh}. 

If one thinks more carefully about the formula Equ.~(\ref{eq:Omega}) and the fact that it works quite universally for any black holes in general relativity, one will see that to properly understand it one cannot ignore gravity, i.e.~one cannot take $G_{N}=0$, otherwise one gets an infinite number for $\Omega_{\text{BH}}$ which is meaningless, and its statistical origin shouldn't rely on the fine details of the gravitational theory such as how the black hole is constructed using strings and branes by different kinds of compactifications \cite{Strominger:1996sh,David:2002wn}.\footnote{See \cite{Strominger:2009aj} for other arguments for the universality of the Bekenstein-Hawking entropy formula.} Furthermore, the deployment of $G_{N}$ in the numerator indicates that to reproduce Equ.~(\ref{eq:Omega}) from a state-counting argument, nonperturbative effects of gravity should be involved. With these understandings in mind, let's briefly revisit the bag of gold paradox and the request for a state-counting derivation of Equ.~(\ref{eq:Omega}). In the line of arguments for the bag of gold paradox one assumes that 1) the backreaction of the excitations to the geometry can be ignored; 2) we have a notion of locality, i.e.~the excitations behind the horizon won't affect the region outside. However, if one considers dynamical gravity, none of these two assumptions are generally valid as even perturbative effects in $G_{N}$ will invalidate them \cite{Donnelly:2017jcd,Donnelly:2018nbv,Laddha:2020kvp,Chowdhury:2020hse,Geng:2021hlu,Chowdhury:2021nxw,Geng:2023zhq}. Nonetheless, to fully resolve the bag of gold paradox one has to show that those naively orthogonal states created by exciting the interior are in fact not orthogonal and the number of truly linearly independent states is upper bounded by Equ.~(\ref{eq:Omega}). Interestingly, this objective is of a state-counting nature and hence should be closely related to the request for a state-counting proof of Equ.~(\ref{eq:Omega}). Moreover, it in fact suggests the way to prove Equ.~(\ref{eq:Omega}), and once Equ.~(\ref{eq:Omega}) is proved the bag of gold paradox should be easily resolved. This way of the proof works as follows that one firstly has to construct an over-complete set of microstates for the black hole, then show that these states are overlapping due to nonperturbative gravitational effects, and finally prove that only $e^{S_{\text{BH}}}$ of them are linearly independent.\footnote{See \cite{papadodimas:2013kwa,essay2020,Chakravarty:2020wdm} for earlier proposals of the potential use of nonperturbatively small overlaps between states to reconcile the Bekenstein-Hawking entropy formula with the seemingly large number of bulk states.} The goal of this paper is to provide such a construction and the proof for single-sided black holes in anti-de Sitter spacetime. We will adapt the methods developed in \cite{Kourkoulou:2017zaj,Goel:2018ubv,Penington:2019kki,Sasieta:2022ksu,Chandra:2022fwi,Balasubramanian:2022gmo,Balasubramanian:2022lnw,Climent:2024trz} to our studies and we will focus on the (2+1)-dimensional case for explicit analytic calculations in our paper. Though our results can be easily generalized to higher spacetime dimensions.

This paper is organized as follows. In Sec.~\ref{sec:singlesidecon} we firstly articulate the concept of the black hole microstates and then we provide an explicit construction of single-sided black hole microstates. In Sec.~\ref{sec:singleE} we provide a new set of black hole microstates and we discuss how one can explicitly reproduce the Bekenstein-Hawking entropy formula for single-sided black holes using this set of microstates. This new set of microstates are motivated by constructing approximates for the microstates we constructed in Sec.~\ref{sec:singlesidecon} which are under better control in relevant calculations. We conclude our paper with discussions in Sec.~\ref{sec:conclusion}. Some technical details are collected in Appendix.~\ref{sec:brane} and Appendix.~\ref{sec:shell}.

\section{Single-sided Black Hole Microstates and the Puzzle}\label{sec:singlesidecon}
In this section, we set up the stage for the question we are studying. We firstly articulate the definition of black hole microstates and then we construct an explicit set of microstates for single-sided black holes in (2+1) spacetime dimensions.

\subsection{Definition of the Black Hole Microstate}\label{sec:definition}
As we discussed in the introduction, black holes behave thermodynamically, so the microstates of a black hole should also behave thermodynamically with the same macroscopic parameters as the black hole. However, from the quantum statistical mechanics point of view, thermodynamical properties are coarse-grained properties of the quantum states. Hence the black hole microstates should have the same coarse-grained properties as a thermal state $\rho_{\text{th}}(\beta)$, where 
\begin{equation}
    \rho_{\text{th}}(\beta)=\frac{e^{-\beta\hat{H}}}{\Tr e^{-\beta \hat{H}}}\,,
\end{equation}
is the thermal density matrix with inverse temperature $\beta$ and $\hat{H}$ is the Hamiltonian operator. Since we are considering black holes in anti-de Sitter spacetime, we can use the AdS/CFT correspondence to discuss black hole microstates more systematically. A microstate $\ket{\Psi(\beta)}$ of a black hole in AdS$_{d+1}$ is dual to a state $\ket{\Psi(\beta)}$ in the dual CFT$_{d}$ that lives on the asymptotic boundary of the black hole spacetime. The state $\ket{\Psi(\beta)}$ has almost the same macroscopic properties as the black hole, for example
\begin{equation}
\bra{\Psi(\beta)}\hat{H}\ket{\Psi(\beta)}\in [M-\Delta E,M+\Delta E]\,,\quad\text{with }\Delta E\ll M\,,\label{eq:exp}
\end{equation}
where $\hat{H}$ is the CFT$_{d}$ Hamiltonian operator and $M$ is the mass of the black hole. Moreover, if one probes the state $\ket{\Psi(\beta)}$ using low-point correlators of simple local operators $\hat{O}(x)$ then one wouldn't be able to distinguish the microstate $\ket{\Psi(\beta)}$ and the thermal state $\rho_{\text{th}}(\beta)$ in the limit of large black hole entropy, i.e.~
\begin{equation}
    \bra{\Psi(\beta)}\hat{O}(x_{1})\cdots\hat{O}(x_{n})\ket{\Psi(\beta)}= \Tr\Big(\rho_{\text{th}}(\beta)\hat{O}(x_{1})\cdots\hat{O}(x_{n})\Big)+\mathcal{O}(\frac{1}{S_{\text{BH}}^{\alpha}})\,,\label{eq:microstatedef}
\end{equation}
with $n\ll S_{\text{BH}}$\footnote{The $n\ll S_{BH}$ can be heuristically understood in the following way. The Hilbert space of black hole microstates should presumably have dimension $e^{S_{\text{BH}}}$. Therefore a density matrix in this Hilbert space has $e^{2S_{\text{BH}}}$ components. One can think of the $n$ points as a lattice with $n$ sites and one qubit on each site. Therefore if one has the control of $n$ such qubits then one is able to decode $2^{n}$ bits of information. Hence if $n$ is of order $S_{\text{BH}}$ then one is for sure able to distinguish between different black hole microstates. See \cite{Ghosh:2017pel} for another illustration of this bound.}, $\alpha$ is a positive real number and for states obeying the \textit{eigenstate thermalization hypothesis} (ETH) the deviation from the thermal correlator has to be $\mathcal{O}(e^{-S_{\text{BH}}})$ \cite{srednicki1994chaos,Lashkari:2016vgj,Raju:2018xue}.  In the AdS/CFT duality, simple operators in CFT$_{d}$ are single-trace operators whose bulk duals are weakly interacting quantum fields.\footnote{Such operators are also called the \textit{generalized free fields} \cite{ElShowk:2011ag}.} Hence if a microstate has the same macroscopic properties as the black hole in the sense of Equ.~(\ref{eq:exp}) and satisfies Equ.~(\ref{eq:microstatedef}) then it is a black hole microstate,\footnote{We note that states obeying Equ.~(\ref{eq:exp}) and Equ.~(\ref{eq:microstatedef}) are also called \textit{typical black hole microstates} in the literature \cite{Raju:2018xue,Raju:2020smc}. This is because states partially violating Equ.~(\ref{eq:exp}) and Equ.~(\ref{eq:microstatedef}) only occupy an exponentially small volume in the  Hilbert space of black hole microstates. In this paper, we don't consider those states partially violating Equ.~(\ref{eq:exp}) and Equ.~(\ref{eq:microstatedef}) as, even though some of them might form a basis of the Hilbert space of black hole microstates, they shouldn't be relevant for a \textit{universal} state-counting derivation of the Bekenstein-Hawking entropy formula.} otherwise it is not.

\subsection{A Family of Black Hole Microstates and the Puzzle}\label{sec:microcons}
\begin{figure}
    \centering
    \subfloat[AdS$_{d+1}$ Eternal Black Hole\label{pic:eternal}]{
    \begin{tikzpicture}[scale=1]
       \draw[-,very thick] 
       decorate[decoration={zigzag,pre=lineto,pre length=5pt,post=lineto,post length=5pt}] {(-2.5,0) to (2.5,0)};
       \draw[-,very thick,black] (-2.5,0) to (-2.5,-5);
       \draw[-,very thick,black] (2.5,0) to (2.5,-5);
         \draw[-,very thick] 
       decorate[decoration={zigzag,pre=lineto,pre length=5pt,post=lineto,post length=5pt}] {(-2.5,-5) to (2.5,-5)};
       \draw[-,very thick] (-2.5,0) to (2.5,-5);
       \draw[-,very thick] (2.5,0) to (-2.5,-5);
    \end{tikzpicture}}
    \hspace{2cm}
     \subfloat[Single-sided AdS$_{d+1}$ Black Hole\label{pic:singlebh}]{
    \begin{tikzpicture}[scale=1]
       \draw[-,very thick] 
       decorate[decoration={zigzag,pre=lineto,pre length=5pt,post=lineto,post length=5pt}] {(-2.5,0) to (2.5,0)};
       \draw[-,very thick,black] (-2.5,0) to (-2.5,-5);
       \draw[-,very thick,black] (2.5,0) to (2.5,-5);
         \draw[-,very thick] 
       decorate[decoration={zigzag,pre=lineto,pre length=5pt,post=lineto,post length=5pt}] {(-2.5,-5) to (2.5,-5)};
       \draw[-,very thick] (-2.5,0) to (2.5,-5);
       \draw[-,very thick] (2.5,0) to (-2.5,-5);
       \draw[-,very thick,blue] (-1.49,0) arc (150:210:5);
       \draw[fill=gray, draw=none, fill opacity = 0.5]  decorate[decoration={zigzag,pre=lineto,pre length=5pt,post=lineto,post length=5pt}] {(-2.5,0) to (-1.49,0)} arc (150:210:5) decorate[decoration={zigzag,pre=lineto,pre length=5pt,post=lineto,post length=5pt}] {(-1.49,-5) to (-2.5,-5)} to (-2.5,0); 
    \end{tikzpicture}}
    \caption{\small\textbf{a)} The Penrose diagram of an eternal AdS$_{d+1}$ black hole spacetime. There are two exteriors and two asymptotic boundaries where the two dual CFT$_{d}$'s live. These two CFT$_{d}$'s are in the TFD state Equ.~(\ref{eq:TFD}). \textbf{b)} The Penrose diagram describing the dual CFT state Equ.~(\ref{eq:Psi}). The state Equ.~(\ref{eq:Psi}) is obtained by projecting out one CFT$_{d}$ in the TFD state Equ.~(\ref{eq:TFD}) and the dual bulk description is that a brane in the black hole spacetime cuts off one asymptotic boundary. The blue curve is the locus of the brane and the grey shaded region is cut off by this brane.}   
\end{figure}
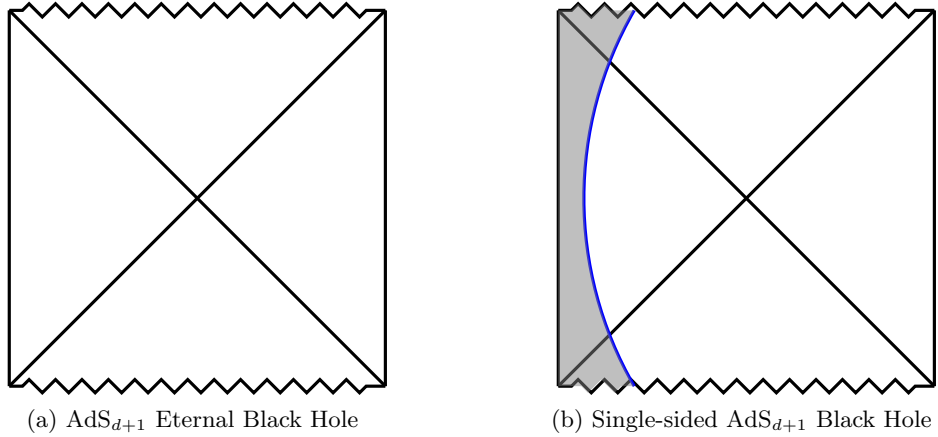

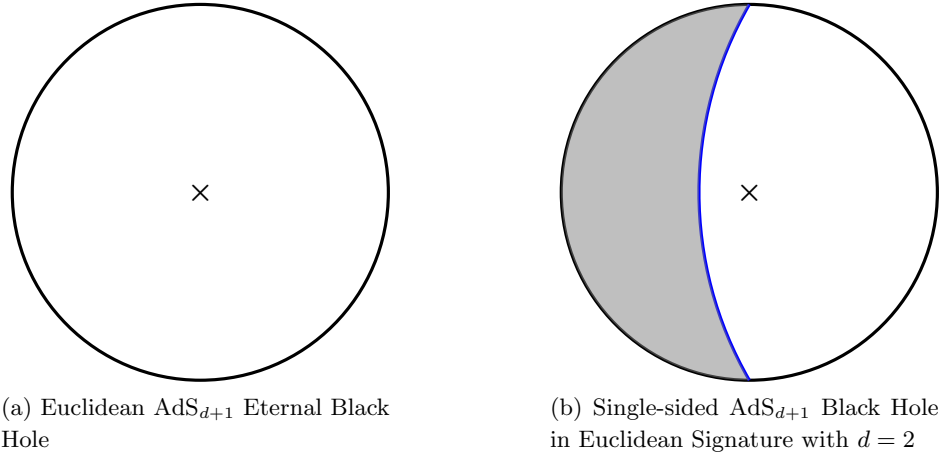
\begin{figure}
\begin{centering}
\subfloat[Euclidean AdS$_{d+1}$ Eternal Black Hole\label{pic:eternalE}]{
    \begin{tikzpicture}[scale=1]
\draw[-,very thick,black] (2.5,0) arc (0:360:2.5);
\node at (0,0) {\textcolor{black}{$\cross$}};
\end{tikzpicture}}
    \hspace{2cm}
     \subfloat[Single-sided AdS$_{d+1}$ Black Hole in Euclidean Signature with $d=2$\label{pic:singlebhE}]{
    \begin{tikzpicture}[scale=1]
    \draw[-,very thick,black] (2.5,0) arc (0:360:2.5);
    \draw[-,very thick,blue] (0,2.5) arc (150:210:5);
     \draw[fill=gray, draw=none, fill opacity = 0.5] (0,2.5) arc (150:210:5) arc (-90:-270:2.5);
\node at (0,0) {\textcolor{black}{$\cross$}};
     \end{tikzpicture}}
\caption{\small\textbf{a)} The Euclidean signature Penrose diagram of an eternal AdS$_{d+1}$ black hole. The radial direction of the disk is the $r$-direction and the polar angle denotes the $\tau$-direction. The boundary of the disk is the asymptotic boundary which supports an Euclidean CFT$_{d}$. \textbf{b)} The brane locus in the Euclidean signature for $d=2$. The brane is denoted by the blue curve and it always subtends half of the domain of the Euclidean time direction. The grey shaded region is cut off. This configuration describes the Euclidean time evolution of the Cardy state $\ket{B_{i}}$ and it computes the CFT quantity $\bra{\Psi_{i}(\beta)}\ket{\Psi_{i}(\beta)}$.}
\label{pic:demoE}
\end{centering}
\end{figure}

It has been pointed out in \cite{Maldacena:2001kr} that an eternal black hole with high enough temperature in $(d+1)$-dimensional anti-de Sitter spacetime (AdS$_{d+1}$) is described by a thermal field double state (TFD) in its dual $d$-dimensional conformal field theory (CFT$_{d}$). The metric of such a black hole is
\begin{equation}
    ds^2=-f(r)dt^2+\frac{dr^2}{f(r)}+r^2 d\Omega_{d-1}^2\,,\quad f(r)=1+r^{2}-\frac{16\pi G_{N}M}{(d-1)\Omega_{d-1}r^{d-2}}\,,\label{eq:BHmetric}
\end{equation}
where the inverse temperature is $\beta=\frac{4\pi r_H}{d r_H^2+d-2}$ and in this paper we focus on the case with $d=2$. This black hole spacetime has two asymptotic boundaries (see Fig.~\ref{pic:eternal}) connected to each other through a bulk Einstein-Rosen bridge. The two asymptotic boundaries support the two dual CFT$_{2}$'s forming the TFD state
\begin{equation}
    \ket{\text{TFD}}=\frac{1}{\sqrt{Z(\beta)}}\sum_{n}e^{-\frac{\beta}{2}E_{n}}\ket{E_{n}^{*}}\ket{E_{n}}\,,\label{eq:TFD}
\end{equation}
where $\beta$ denotes the inverse temperature of the black hole, $E_{n}$'s are the energy eigenvalues of the CFT$_{2}$ on the circle, $Z(\beta)=\sum_{n}e^{-\beta E_{n}}$ is the thermal partition function and $\ket{E_{n}^{*}}$ denotes $\hat{\Theta}\ket{E_{n}}$ with $\hat{\Theta}$ being the $CPT$ conjugation \cite{Harlow:2014yka}. To construct a one-sided black hole microstate, one can project one of the CFT$_{2}$'s in the $\ket{\text{TFD}}$ state onto a smeared Cardy state $e^{-\frac{\beta}{4}\hat{H}}\ket{B}$ and this gives
\begin{equation}
\begin{split}
    \ket{\Psi_{B}^{\beta}}&=\sum_{n}e^{-\frac{\beta}{2}E_{n}}\bra{B}e^{-\frac{\beta}{4}\hat{H}}\ket{E_{n}^{*}}\ket{E_{n}}\\&=\sum_{n}e^{-\frac{\beta}{4}E_{n}}\bra{E_{n}}\ket{B}\ket{E_{n}}\\&=\sum_{n}e^{-\frac{\beta}{4}E_{n}}\ket{E_{n}}\bra{E_{n}}\ket{B}\\&=e^{-\frac{\beta}{4}\hat{H}}\ket{B}\,,\label{eq:Psi}
    \end{split}
\end{equation}
where $\hat{H}$ denotes the Hamiltonian of the $t$-evolution which is also the Hamiltonian of the CFT$_{2}$ on the circle, we have removed the over factor $\frac{1}{\sqrt{Z(\beta)}}$ for simplicity, and we have used the fact that in the Euclidean signature $CPT$ maps $\hat{H}$ to $-\hat{H}$ and it is anti-unitary. This projection removes the entanglement between the two CFT$_{2}$'s, so one expects that the effect in the bulk black hole spacetime is to cut off one of the two asymptotic regions, and how much it cuts off depends on the state $\ket{B}$ (see Fig.~\ref{pic:singlebh}). The surface that cuts off the projected asymptotic boundary is a brane (the blue curve of Fig.~\ref{pic:singlebh}), and the best way to understand this surface is to go to the Euclidean signature in the time-direction $t\rightarrow-i\tau$. In the Euclidean signature, the bulk geometry Equ.~(\ref{eq:BHmetric}) is a disk in the $r-\tau$ directions (see Fig.~\ref{pic:eternalE}). From the boundary point of view, the state $\ket{B}$ simply creates a boundary on the $\tau$ direction, and since it is a Cardy state this boundary is a conformal boundary \cite{Cardy:2004hm}. The bulk dual of a conformal boundary that corresponds to a Cardy state is a Karch-Randall brane \cite{Karch:2000ct}. Hence we can see that the brane that cuts off the projected asymptotic boundary is a Karch-Randall brane. After the geometry of the brane is understood in the Euclidean signature, we can analytically continue back to the Lorenzian signature. The gravitational system with a Karch-Randall brane is described by the action
\begin{equation}
S=-\frac{1}{16\pi G_{N}}\int_{\mathcal{M}_{3}} d^{3}x\sqrt{-g} (R-2\Lambda)-\frac{1}{8\pi G_{N}}\int_{\mathcal{B}_{2}}d^{3}x\sqrt{-h}(K-T)-\frac{1}{8\pi G_{N}}\int_{\partial\mathcal{M}_{3}}d^{2}x\sqrt{-h}K\,,\label{eq:action1}
\end{equation}
where $\mathcal{M}_{3}$ denotes the bulk with asymptotic boundary $\partial\mathcal{M}_{2}$, $\mathcal{B}_{2}$ denotes the brane, $K$ is the trace of the extrinsic curvature of the corresponding hypersurface for the brane and $T$ is the brane tension. This action gives two sets of equations of motion
\begin{equation}
    R_{\mu\nu}-\frac{1}{2}g_{\mu\nu} R-\Lambda g_{\mu\nu}=0\,,\quad K h_{ab}- K_{ab}=T h_{ab}\,,\label{eq:eom1}
\end{equation}
where the first is the bulk Einstein's field equation and the second is the brane embedding equation. The Einstein's field equation is solved by the bulk geometry Equ.~(\ref{eq:BHmetric}) and the brane embedding equation determines the geometric configuration of the brane in the bulk. The brane embedding equation in the bulk spacetime Equ.~(\ref{eq:BHmetric}) has been analyzed in detail in the Euclidean signature in Appendix.~\ref{sec:brane}. The result is shown in Fig.~\ref{sec:singleE}. The branes we consider are spherically symmetric, i.e.~they wrap the bulk $S^{1}$ sector. The brane always lies behind the black hole horizon and subtends an interval of length $\frac{\beta}{2}$ in the Euclidean time direction where $\beta$ is the inverse of the bulk black hole temperature and it is the same  $\beta$ as in Equ.~(\ref{eq:Psi}). We only consider positive tension branes for which the bulk region behind the brane is cut off and the larger the brane tension is the smaller the cutoff region is. The brane tension takes value in $[0,1)$. When the brane tension is zero, the brane goes through the horizon straightly, cutting off exactly a half of the bulk spacetime. As the brane tension approaches one, the brane is approaching the cutoff asymptotic boundary.\footnote{That is the gray shaded region in Fig.~\ref{pic:singlebhE} is exactly half of the whole diagram if $T=0$ and it gets smaller if the tension $T$ is closer to one from below.} Hence in the Lorentzian signature, the brane is always behind the bulk black hole horizon (see Fig.~\ref{pic:singlebh}). 

\begin{figure}
    \centering
    \subfloat[AdS$_{d+1}$ Eternal Black Hole\label{pic:eternalcorrelator}]{
    \begin{tikzpicture}[scale=1]
       \draw[-,very thick] 
       decorate[decoration={zigzag,pre=lineto,pre length=5pt,post=lineto,post length=5pt}] {(-2.5,0) to (2.5,0)};
       \draw[-,very thick,black] (-2.5,0) to (-2.5,-5);
       \draw[-,very thick,black] (2.5,0) to (2.5,-5);
         \draw[-,very thick] 
       decorate[decoration={zigzag,pre=lineto,pre length=5pt,post=lineto,post length=5pt}] {(-2.5,-5) to (2.5,-5)};
       \draw[-,very thick] (-2.5,0) to (2.5,-5);
       \draw[-,very thick] (2.5,0) to (-2.5,-5);
       \node at (2.5,-2.5) {\textcolor{red}{$\bullet$}};
    \end{tikzpicture}}
    \hspace{2cm}
     \subfloat[Single-sided AdS$_{d+1}$ Black Hole\label{pic:singlebhcorrelator}]{
    \begin{tikzpicture}[scale=1]
       \draw[-,very thick] 
       decorate[decoration={zigzag,pre=lineto,pre length=5pt,post=lineto,post length=5pt}] {(-2.5,0) to (2.5,0)};
       \draw[-,very thick,black] (-2.5,0) to (-2.5,-5);
       \draw[-,very thick,black] (2.5,0) to (2.5,-5);
         \draw[-,very thick] 
       decorate[decoration={zigzag,pre=lineto,pre length=5pt,post=lineto,post length=5pt}] {(-2.5,-5) to (2.5,-5)};
       \draw[-,very thick] (-2.5,0) to (2.5,-5);
       \draw[-,very thick] (2.5,0) to (-2.5,-5);
       \draw[-,very thick,blue] (-1.49,0) arc (150:210:5);
       \draw[fill=gray, draw=none, fill opacity = 0.5]  decorate[decoration={zigzag,pre=lineto,pre length=5pt,post=lineto,post length=5pt}] {(-2.5,0) to (-1.49,0)} arc (150:210:5) decorate[decoration={zigzag,pre=lineto,pre length=5pt,post=lineto,post length=5pt}] {(-1.49,-5) to (-2.5,-5)} to (-2.5,0); 
       \draw[-,very thick] (2.5,0) to (-2.5,-5);
       \node at (2.5,-2.5) {\textcolor{red}{$\bullet$}};
       \draw[-,thick,red] (2.5,-2.5) to (-2.15,-2.5); 
    \end{tikzpicture}}
    \caption{\small The red dot on the asymptotic boundary denotes the operator insertion. \textbf{a)} Such a one-point function is zero in the TFD state Equ.~(\ref{eq:Psi}). \textbf{b)} Such a one-point function is generally nonzero in the states Equ.~(\ref{eq:Psi}). The red line connecting the boundary operator insertion and the brane is the geodesic that computes this one-point function in the WKB regime following Equ.~(\ref{eq:WKB}).}   
\end{figure}
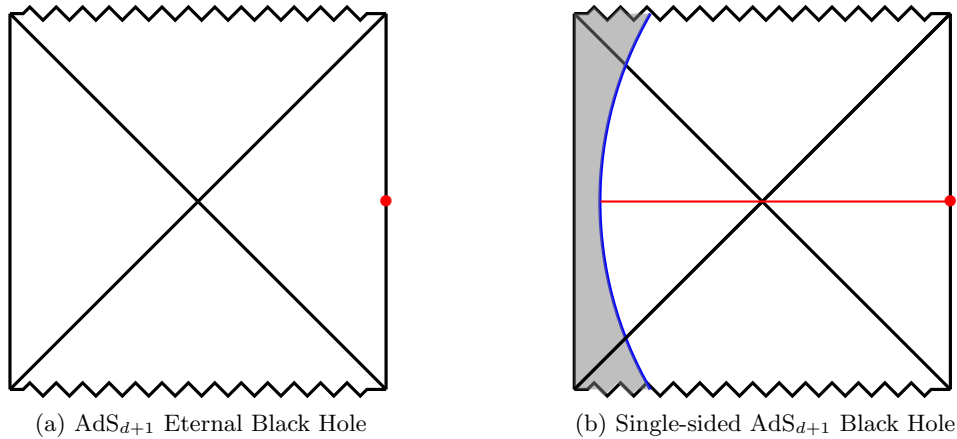

However, we should note that the states $\ket{\Psi_{B}^{\beta}}$ with different brane tensions but the same $\beta$ are not generally microstates of the single-sided black hole with inverse temperature $\beta$. This is because black hole microstates have to satisfy Equ.~(\ref{eq:microstatedef}) which a general state $\ket{\Psi_{B}^{\beta}}$ doesn't satisfy. This can be seen as follows. The thermal state $\rho_{\text{th}}(\beta)$ can be obtained from the thermal field double state by tracing out one copy of the CFT$_{d}$, i.e.
\begin{equation}
    \rho_{\text{th}}(\beta)=\Tr_{\mathcal{H}_{L}} \ket{\text{TFD}}\bra{\text{TFD}}\,,
\end{equation}
where $\mathcal{H}_{L}$ denotes the Hilbert space of the CFT$_{d}$ that lives on the left asymptotic boundary of the eternal AdS$_{d+1}$ black hole. Therefore one can think of $\rho_{\text{th}}(\beta)$ as only describing the CFT$_{d}$ that lives on the right asymptotic boundary of the black hole (i.e.~the right asymptotic boundary of Fig.~\ref{pic:eternal}), and the correlators of simple operators in the state $\rho_{\text{th}}(\beta)$ can be computed using the bulk dual as operators inserted on the right asymptotic boundary (see Fig.~\ref{pic:eternalcorrelator}). Similarly one can compute the correlators of simple operators in the state $\ket{\Psi_{B}^{\beta}}$ using the bulk dual as operators inserted on the asymptotic boundary (see Fig.~\ref{pic:singlebhcorrelator}). In fact, one can see that it is easy to probe the brane using simple operators of the CFT$_{d}$ and even one-point functions can do the job. For example, one can consider the one-point function of a CFT$_{d}$ scalar primary $\hat{O}_{\Delta}(x)$ which is dual to a heavy scalar field $\phi(x)$ in the bulk for which one can compute $\Tr (\rho_{\beta}\hat{O}_{\Delta}(x))$ and $\bra{\Psi_{B}^{\beta}}\hat{O}_{\Delta}(x)\ket{\Psi_{B}^{\beta}}$ using the WKB approximation. The WKB approximation gives us
\begin{equation}
    \Tr (\rho_{\text{th}}(\beta)\hat{O}_{\Delta}(x))=0\,,
\end{equation}
and 
\begin{equation}
    \bra{\Psi_{B}^{\beta}}\hat{O}_{\Delta}(x)\ket{\Psi_{B}^{\beta}}=e^{-m_{\Delta}l^{\text{ren}}}\,,\label{eq:WKB}
\end{equation}
where $l^{\text{ren}}$ is the renormalized geodesic length between the boundary operator insertion and the brane. For $d=2$ and a zero operator insertion time we have
\begin{equation}
    \begin{split}
l^{\text{ren}}&=\int_{r_{H}}^{r_{\epsilon}}\frac{dr}{\sqrt{r^{2}-r_{H}^{2}}}+\int_{r_{H}}^{r_{c}}\frac{dr}{\sqrt{r^{2}-r_{H}^{2}}}-\log r_{\epsilon}\,,\\&=\log\sqrt{\frac{1+T}{1-T}}-\log \frac{r_{H}}{2}\,,
    \end{split}
\end{equation}
where $r_{c}$ is the turning point of the brane which is given in Equ.~(\ref{eq:braneconstraint}) and we have taken the $r_{\epsilon}\rightarrow\infty$ limit. Thus we can see that the state $\ket{\Psi_{B}^{\beta}}$ is a black hole microstate if and only if $T$ is close to 1, and from the bulk perspective this means that the brane is close enough to the cutoff asymptotic boundary. As a result, we are able to construct a genuine set of black hole microstates $\ket{\Psi_{B}^{\beta}}$ with $T$ close to 1, and this is a continuous one-parameter set of black hole microstates.

Nevertheless, the puzzle comes if one thinks of the statistical interpretation of the Bekenstein-Hawking entropy formula Equ.~(\ref{eq:BH}). This is because the above states $\ket{\Psi_{B}^{\beta}}$ with equal $\beta$ comprise a continuous one-parameter set of microstates for the single-sided black hole with inverse Hawking temperature $\beta$, and the potential of this continuous set is clearly larger than the exponential of the Bekenstein-Hawking entropy. Thus to reconcile these two perspectives and give a state-counting derivation of the Bekenstein-Hawking entropy formula, one has to show that these microstates $\ket{\Psi_{B}^{\beta}}$ have nonperturbatively small overlaps which gives exactly $e^{S_{\text{BH}}}$ as the number of linearly independent states among them.

\section{Microscopic Derivation of the Entropy of the Singled-sided Black Hole}\label{sec:singleE}
As we have discussed, to resolve the puzzle we proposed in Sec.~\ref{sec:singlesidecon} using the bulk gravitational theory, we firstly have to compute the overlap between the different states $\ket{\Psi_{B}^{\beta}}$ with the same $\beta$. For the sake of simplicity, let's denote the state $\ket{\Psi_{B_{i}}^{\beta}}$ as $\ket{\Psi_{i}(\beta)}$ and we want to compute the overlap
\begin{equation}
    G_{ij}^{\Psi}=\frac{\bra{\Psi_{i}(\beta)}\ket{\Psi_{j}(\beta)}}{\sqrt{\bra{\Psi_{i}(\beta)}\ket{\Psi_{i}(\beta)}}\sqrt{\bra{\Psi_{j}(\beta)}\ket{\Psi_{j}(\beta)}}}\,,
\end{equation}
using the bulk gravitational theory. We consider the regime where $G_{N}$ is small such that gravitational path integrals can be computed using the saddle point approximation.\footnote{Notice that as we discussed the black hole entropy formula Equ.~(\ref{eq:BH}) is in fact only true to leading order in $G_{N}$, if one takes $G_{N}$ as the bare Newton's constant, as there are subleading corrections to it in higher orders of $G_{N}$. Hence to understand Equ.~(\ref{eq:BH}), it is enough to only consider leading effect in $G_{N}$, which is captured by the saddle-point approximation and one may replace $G_{N}$ by the renormalized Newton's constant at the end \cite{Susskind:1993if}.} 

The normalization factor $\bra{\Psi_{i}(\beta)}\ket{\Psi_{i}(\beta)}$ gets contribution from the configuration as depicted in Fig.~\ref{pic:singlebhE}, where a single Karch-Randall brane sits behind the black hole horizon and connects the two places on the boundary where the states $\ket{B_{i}}$ and $\bra{B_{i}}$ are created. The saddle-point approximation tells us that $\bra{\Psi_{i}(\beta)}\ket{\Psi_{i}(\beta)}=e^{-S_{i}^{\text{ren}}}$, where $S_{i}^{\text{ren}}$ is the renormalized action, i.e.~Equ.~(\ref{eq:action1}) together with an appropriate counter term (see Appendix.~\ref{sec:brane} for details), evaluated on the configuration like Fig.~\ref{pic:singlebhE} in the Euclidean signature. Since this is the same as the rule of evaluating the partition function using gravitational path integral, let's denote $e^{-S_{i}^{\text{ren}}}$ by $Z_{i}[\beta]$. We have computed in detail in Appendix.~\ref{sec:brane} that $Z_{i}[\beta]$ is in fact independent of $i$ and it is given by
\begin{equation}
    Z_{i}[\beta]=\sqrt{Z[\beta]}\,,\quad\forall i\,,\label{eq:Zi}
\end{equation}
where $Z[\beta]$ is the Gibbons-Hawking partition function of an AdS$_{3}$ BTZ black hole with inverse temperature $\beta$.

Nevertheless, it is unclear what kind of bulk configurations will contribute to $\bra{\Psi_{i}(\beta)}\ket{\Psi_{j}(\beta)}$ for $i\neq j$. This is because the two states $\ket{\Psi_{i}(\beta)}$ and $\ket{\Psi_{j}(\beta)}$ require two branes with different tensions so it is not clear how to embed these configurations smoothly in a single bulk. Presumably the two branes will intersect deep in the bulk \cite{Geng:2021iyq} and a phase transition happens at that intersection point for the brane to transit from being with tension $T_{i}$ to $T_{j}$. Moreover, since this phase transition happens deep in the bulk, the detailed mechanism of the transition is important to correctly evaluate the overlap $\bra{\Psi_{i}(\beta)}\ket{\Psi_{j}(\beta)}$ which however will significantly complicate our calculation. 

\subsection{A New Set of Microstates}\label{sec:approxstates}
To avoid the above complication of the brane intersection, we will instead consider a new set of states $\{\ket{i}\}$ and the higher moments of $G_{ij}$ associated to them. This new set of microstates is motivated by an attempt to properly approximate the states $\ket{\Psi_{i}(\beta)}$ but they are equally well as a set of black hole microstates. In this subsection, we discuss this new set of microstates and in the next two subsections we consider the higher moments of $G_{ij}$ for this new set of states and explain how we can extract $\Omega_{\text{BH}}$ from them. 

Let's denote the states $\ket{\Psi_{B_{0}}^{\beta'_{i}}}$ corresponding to a brane with reference tension $T_{0}$ in a black hole with inverse temperature $\beta'_{i}$ as $\ket{\Psi_{0}^{i}}$ and let's consider the states
\begin{equation}
    \ket{i}=e^{-\frac{\beta_{i}}{2}\hat{H}}\hat{O}_{i}e^{\frac{\beta_{i}}{4}\hat{H}}\ket{\Psi_{0}^{i}}\,.
\end{equation}
The CFT$_{2}$ operator $\hat{O}_{i}$ is supported on the whole $S^{1}$ on which the CFT$_{2}$ lives and it is constructed such that its bulk dual is a shell of pressureless dust with mass $m_{i}$ \cite{Anous:2016kss}. The full gravitational action with the shell included is given by
\begin{equation}
\begin{split}
S=&-\frac{1}{16\pi G_{N}}\int_{\mathcal{M}_{3}} d^{3}x\sqrt{-g} (R-2\Lambda)-\frac{1}{8\pi G_{N}}\int_{\mathcal{B}_{2}}d^{3}x\sqrt{-h}(K-T)\\&+\int_{\mathcal{S}}d^{2}x\sqrt{-h}\sigma_{i}-\frac{1}{8\pi G_{N}}\int_{\partial\mathcal{M}_{3}}d^{2}x\sqrt{-h}K\,,\label{eq:action2}
\end{split}
\end{equation}
where we followed the same notations as in Equ.~(\ref{eq:action1}), $\mathcal{S}$ denotes the shell and $\sigma_{i}$ is the mass density of the shell. The mass of the shell is conserved so we have
\begin{equation}
    \sigma_{i}(s)V(s)=m_{i}\,,\label{eq:shellconservation}
\end{equation}
where $s$ denotes the proper time along the shell and $V(s)=\int dx \sqrt{h_{s}}$, for which $h_{s,ab}=h_{ab}+u_{a}u_{b}$ is the spatial metric of the shell with $u_{a}$ the unitly normalized velocity vector of the shell, is the spatial volume of the shell. The above action Equ.~(\ref{eq:action2}) gives three sets of equations-- the bulk Einstein's field equation, the brane embedding equation and the junction condition through the shell. The first two equations are the same as those in Equ.~(\ref{eq:eom1}) and in the Euclidean signature the junction condition through the shell reads
\begin{equation}
    h_{ab}\Delta K-\Delta K_{ab}=8\pi G_{N}\sigma_{i} u_{a}u_{b}\,,\label{eq:shelleom}
\end{equation}
where $\Delta K_{ab}$ is the jump of the extrinsic curvature of the shell when we go across it and we have used Equ.~(\ref{eq:shellconservation}) to get Equ.~(\ref{eq:shelleom}) from the variational calculation of Equ.~(\ref{eq:action2}). We only consider spherically symmetric configurations, i.e.~the shell will wrap the $S^{d-1}$ which is the same as what the branes will do. The backreacted geometry from the shell is a sewing of two black hole spacetimes with different mass. We will use $M_{+}$ to denote the mass of the black hole spacetime which contains large portions of the asymptotic boundary and $M_{-}$ to denote the black hole spacetime which contains small portions of the asymptotic boundary (see Appendix.~\ref{sec:shell} for details).\footnote{Thus the large portion of the asymptotic boundary describes the state that is already excited by the shell.}

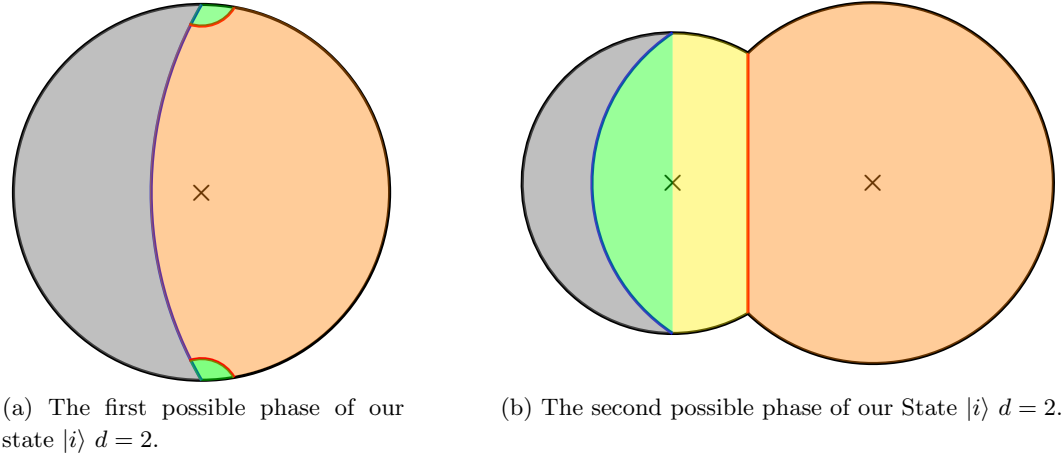
\begin{figure}
\begin{centering}
     \subfloat[The first possible phase of our state $\ket{i}$ $d=2$\label{pic:statei1}.]{
    \begin{tikzpicture}[scale=1]
    \draw[-,very thick,black] (2.5,0) arc (0:360:2.5);
    \draw[-,very thick,blue] (0,2.5) arc (150:210:5);
     \draw[fill=gray, draw=none, fill opacity = 0.5] (0,2.5) arc (150:210:5) arc (-90:-270:2.5);
     \draw[-,very thick,red] (0.434,-2.462) arc (30:109:0.5);
     \draw[-,very thick,red] (0.434,2.462) arc (-30:-109:0.5);
\node at (0,0) {\textcolor{black}{$\cross$}};
\draw[fill=green, draw=none, fill opacity = 0.5](0.434,-2.462) arc (30:109:0.5) arc (206.5:210:5) arc (-90:-80:2.5);
\draw[fill=green, draw=none, fill opacity = 0.4](0.434,2.462) arc (-30:-109:0.5) arc (153.5:150:5) arc (90:80:2.5);
\draw[fill=orange, draw=none, fill opacity = 0.4](0.434,2.462) arc (-30:-109:0.5) arc (153.5:206.5:5) arc (109:30:0.5) arc (-80:80:2.5);
     \end{tikzpicture}}
\hspace{1cm}
     \subfloat[The second possible phase of our State $\ket{i}$ $d=2$.\label{pic:statei2}]{
    \begin{tikzpicture}[scale=0.8]
    \draw[-,very thick,black] (1.25,2.165) arc (60:300:2.5);
    \draw[-,very thick,red] (1.25,2.165) to (1.25,-2.165);
\node at (0,0) {\textcolor{black}{$\cross$}};
  \draw[-,very thick,black] (1.25,2.165) arc (180-46.192:-180+46.192:3);
  \node at (1.25+2.077,0) {\textcolor{black}{$\cross$}};
  \draw[-,very thick,blue] (0,2.5) arc (180-56.443:180+56.443:3);
  \draw[fill=gray, draw=none, fill opacity = 0.5] (0,2.5) arc (180-56.443:180+56.443:3) arc (-90:-270:2.5);
  \draw[fill=green, draw=none, fill opacity = 0.4] (0,2.5) arc (180-56.443:180+56.443:3) to (0,2.5);
  \draw[fill=yellow, draw=none, fill opacity = 0.4] (0,2.5) to (0,-2.5) arc (270:300:2.5) to (1.25,2.165) arc (60:90:2.5);
  \draw[fill=orange, draw=none, fill opacity = 0.4] (1.25,2.165) arc (180-46.192:-180+46.192:3) to (1.25,2.165);
     \end{tikzpicture}}
\caption{\small \textbf{a)} The configuration for the first possible phase of the state $\ket{i}$. More precisely, this configuration computes the norm $\bra{i}\ket{i}$. The trajectories of the shell cross the trajectory of the brane suggesting as before that the shell is absorbed by the brane and excites the brane to the tension $T_{i}$. To avoid considering the detailed interaction between the shell and brane, one wants to defer this absorbing process to the deep UV regime. Therefore, one has to make sure that the green region is suppressed and then only the orange region contribute to the on-shell action computing the norm $\bra{i}\ket{i}$. \textbf{b)} The configuration for the second possible phase of the state $\ket{i}$. As in the first phase, this configuration in fact computes the norm $\bra{i}\ket{i}$. }
\label{pic:demoE}
\end{centering}
\end{figure}

The motivation for the construction of the new set of states $\{\ket{i}\}$ is to properly approximate the states $\ket{\Psi_{i}(\beta)}$. This is because one can think from the dual conformal field theory perspective that the shell created by the operator $\hat{O}_{i}$ excites the boundary state $\ket{B_{0}}$, and after the $e^{-\frac{\beta_{i}}{2}\hat{H}}$ smearing it may approximate the state $\ket{\Psi_{i}(\beta)}$ which is associated with a brane of tension $T_{i}>0$. We will see that this approximation indeed works for branes with a large tension for which the shell is heavy but we'd better think of the states $\ket{i}$ as a new set of microstates. To properly excite the boundary state $\ket{B_{0}}$, we want the operator $\hat{O}_{i}$ to be inserted close enough to the reference Cardy state $\ket{B_{0}}$ so we will take $\beta'_{i}\rightarrow\beta_{i}$. In the computation of $\bra{i}\ket{i}$ using the saddle-point approximation of the bulk gravitational path integral, there are two possible phases of saddles as depicted in Fig.~\ref{pic:statei1} and Fig.~\ref{pic:statei2}. With $T_{0}$, $\beta_{i}$, $\Delta\beta=\beta'_{i}-\beta_{i}$ and $m_{i}$ properly chosen, both of the two phases make sense as approximations of the the microstate $\ket{\Psi_{i}({\beta})}$. 

In the first phase, one can think of the shell to be absorbed by the brane such that the brane transits from with tension $T_{0}$ to $T_{i}$. To avoid addressing the detailed phase transition mechanism, we have to make sure that the shell is absorbed by the brane from very early on, i.e.~$\beta'_{i}\approx\beta_{i}\approx\frac{\beta}{2}$. This is ensured in the limit where $\Delta\beta$ is small and $m$ is large such that
\begin{equation}
    \frac{\Delta\beta}{4}\ll\Delta\tau=\frac{\beta}{\pi}\arcsin\frac{\pi}{G_{N}m\beta}\,,
\end{equation}
where we have taken $\beta'_{i}=\beta_{i}=\frac{\beta}{2}$ and we have used Equ.~(\ref{eq:shellconstraint}) for $\Delta\tau$ which is the lifetime of the shell in the bulk. This limit also ensures that in the first phase the shell will cross the brane soon after it is created so the contribution to the partition function from the shell and the green regions in Fig.~\ref{pic:statei1} can be ignored. Therefore the partition function of the first possible phase can be approximated by $\bra{\Psi_{i}(2\beta_{i})}\ket{\Psi_{i}(2\beta_{i})}$
\begin{equation}
    Z_{i}^{1}=\bra{\Psi_{i}(2\beta_{i})}\ket{\Psi_{i}(2\beta_{i})}=Z_{i}(2\beta_{i})=\sqrt{Z[2\beta_{i}]}\,,\label{eq:Z1}
\end{equation}
where we have used Equ.~(\ref{eq:Zi}) and $\beta_{i}\approx\frac{\beta}{2}$. 

In the second phase, one can think in the following way that when the brane and the shell are very close to each other, from the point of view of the region far from them they are effectively fused into a single brane. To make this picture concrete, we again need $\beta'_{i}\rightarrow\beta_{i}$, i.e.~$\Delta\beta$ is small to make sure that the fusion nicely converges to a single state which approximate $\ket{B_{i}}$. Translating into the bulk, we require that there is little room between the shell and the brane in the bulk. This requirement is achieved with a tensionless brane (i.e.~the green region in Fig.~\ref{pic:statei2} degenerates) and it further requires that $\Delta\beta$ is much less than the time the brane spends in the black hole spacetime with mass $M_{-}$, i.e.~the yellow region becomes empty due to the squeezing of the brane and the shell. Therefore we have
\begin{equation}
    \frac{\Delta\beta}{2}\ll \Delta\tau_{-}=\frac{\beta'_{i}}{\pi}\arcsin\frac{2\pi}{\beta'_{i} r_{c}}\,,
\end{equation}
where we have used the fact that the inverse temperature of the black hole spacetime with mass $M_{-}$ is $\beta'_{i}$, i.e.~the spacetime region before the insertion of the shell which is dual to the reference state $\ket{\Psi_{0}^{i}}$, and $r_{c}$ is given in Equ.~(\ref{eq:shellconstraint}). Moreover, to approximate the microstate $\ket{\Psi_{i}}$ we also require: 1) the inverse temperature of the black hole with mass $M_{+}$ to be the same as the black hole for the state $\ket{\Psi_{i}}$; 2) the shell trajectory approximates that of the brane in the state $\ket{\Psi_{i}}$. The first requirement imposes that $\beta_{i}=\beta$, which we note is of a factor two different from that in the first phase. The second requirement can be satisfied if we consider the regime of large shell mass and large brane tension ($T_{i}\rightarrow1$), for which both the shell and the brane are far from the horizon, and close to the behind horizon asymptotic boundary of the $M_{+}$ black hole. In the large shell mass limit we have $r_{c}=2G_{N}m$. With the above requirements satisfied the partition function in the second phase is given by
\begin{equation}
Z^{2}_{i}=Z_{\text{shell}}Z[\beta_{i}]=e^{2m\log2G_{N}m}Z[\beta_{i}]\,,\label{eq:Z2}
\end{equation}
i.e.~the green and yellow regions in Fig.~\ref{pic:statei2} don't contribute, where we used Equ.~(\ref{eq:Sshellren}). 

In summary, to approximate the microstate $\ket{\Psi_{i}}$ in the second phase we have similar requirements as those in the first phase that we need the shell mass to be large and $\Delta\beta$ to be small, i.e.~we parametrically have
\begin{equation}
    \Delta\beta\ll \beta\arcsin\frac{\pi}{G_{N}m\beta}\,,
\end{equation}
and in the large brane tension regime $T_{i}\rightarrow 1$ both phases are able to approximate the microstate $\ket{\Psi_{i}}$. Nevertheless since we haven't carefully normalize any state, to use the state $\ket{i}$ to approximate $\ket{\Psi_{i}}$ in the later calculations we have to know the norm $\bra{i}\ket{i}$. The norm is computed by the dominate partition function among the partition functions of the first phase and the second phase. The partition functions of the two phases are given respectively by Equ.~(\ref{eq:Z1}) and Equ.~(\ref{eq:Z2}) in which $Z[\beta]$ is the high temperature partition function of a black hole with inverse temperature $\beta$. It can be computed following the method in Appendix.~\ref{sec:brane} as
\begin{equation}
    Z[\beta]=e^{\frac{\pi^2}{2\beta G_{N}}}\,.
\end{equation}
Thus, in the high temperature regime $\beta_{i}\ll 1$ we have $Z^{2}_{i}>Z^{1}_{i}$ so the second phase will dominate over the first phase and we have
\begin{equation}
\bra{i}\ket{i}=Z^{2}_{i}=e^{2m\log2G_{N}m}Z[\beta]\,,\label{eq:ii}
\end{equation}
where we have used $\beta_{i}=\beta$.

As a result, we are able to approximate the states $\ket{\Psi_{i}}$ with a large brane tension using states $\ket{i}$ with a tensionless reference brane, heavy shell and $\beta'_{i}\approx\beta_{i}\approx\beta$. The inner product $\bra{i}\ket{i}$ is computed using the gravitational path integral with Fig.~\ref{pic:statei2} as the dominant saddle and Equ.~(\ref{eq:ii}) as the result. Interestingly, as we have discussed in Sec.~\ref{sec:microcons}, this is exactly the regime where the states $\ket{\Psi_{i}(\beta)}$ are black hole microstates. We will use this new set of states $\{\ket{i}\}$ to count the dimension of the Hilbert space of the black hole microstates.

\subsection{Overlaps between the Microstates}\label{sec:overlap}
In this section, we will show how one can compute the higher moments of the overlap matrix
\begin{equation}
    G_{ij}=\frac{\bra{i}\ket{j}}{\sqrt{\bra{i}\ket{i}}\sqrt{\bra{j}\ket{j}}}\,.
\end{equation}
We will discuss how the Bekenstein-Hawking entropy formula Equ.~(\ref{eq:BH}) can be extracted using these higher moments in the next subsection. As we have discussed in Sec.~\ref{sec:microcons} and Sec.~\ref{sec:approxstates}, the states $\ket{i}$ consist of a continuous set of states. So to talk about $G_{ij}$ as a matrix we will take $\Omega$ states out of the set $\{\ket{i}\}$ and study this discrete subset of states. We will see that the number of linearly independent states is in fact independent of $\Omega$ as long as it is large enough.\footnote{Putting in a more speculative way, one can think that at the more fundamental level the shell consists of discrete number of particles, so one may not be bothered by this continuity \cite{Balasubramanian:2022gmo}.}

\begin{figure}
\begin{centering}
\begin{tikzpicture}[scale=1,decoration=snake]
%%bdy1
\draw[-,very thick,black] (0.521,2.954) arc (80:40:3);
\draw[-,very thick,black]  (2.819,1.026) arc (20:-20:3);
\draw[-,very thick,black] (2.298,-1.928) arc (-40:-80:3);
\draw[-,very thick,black] (-0.521,-2.954) arc (-100:-140:3);
\draw[-,very thick,black] (-2.819,-1.026) arc (-160:-200:3);
\draw[-,very thick,black] (-2.298,1.928) arc (-220:-260:3);
%%shell
\draw[-,thick,red] (-0.521,2.954) arc (-180+46.34:-46.34:0.755);
\draw[-,thick,red] (2.819,1.026) arc (-106:-106-87.32:0.755);
\draw[-,thick,red] (2.298,-1.928) arc (106+87.32:106:0.755);
\draw[-,thick,red] (-0.521,-2.954) arc (180-46.34:46.34:0.755);
\draw[-,thick,red] (-2.819,-1.026) arc (74:74-87.32:0.755);
\draw[-,thick,red] (-2.298,1.928) arc (-74+87.32:-74:0.755);
%%bdy2
\draw[-,very thick,black] (0.521,2.954) arc (-33.59:-33.59+247.18:0.625);
\draw[-,very thick,black]  (2.819,1.026) arc (-93:-93+247.18:0.625);
\draw[-,very thick,black] (2.298,-1.928) arc (93-247.18:93:0.625);
\draw[-,very thick,black] (-0.521,-2.954) arc (180-33.59:(180-33.69)+247.18:0.625);
\draw[-,very thick,black] (-2.819,-1.026) arc (87:87+247.18:0.625);
\draw[-,very thick,black] (-2.298,1.928) arc (-87-247.18:-87:0.625);
%%branes and shades top and bottom
\draw[-,thick,blue] (-0.625,3.3) to (0.625,3.3);
\draw[fill=gray, draw=none, fill opacity = 0.5] (-0.625,3.3) arc (180:0:0.625) to (-0.625,3.3);
\draw[-,thick,blue] (0.625,-3.3) to (-0.625,-3.3);
\draw[fill=gray, draw=none, fill opacity = 0.5](-0.625,-3.3) arc (180:360:0.625) to (-0.625,-3.3);

%%branes and shades others from rightup clockwise
\draw[-,thick,blue] (2.858 +0.625/2, 3.3/2 -0.625*0.866) to (2.858 -0.625/2, 3.3/2 +0.625*0.866);
\draw[fill=gray, draw=none, fill opacity = 0.5] (2.858 +0.625/2, 3.3/2 -0.625*0.866) arc (-60:120:0.625) to (2.858 +0.625/2, 3.3/2 -0.625*0.866);
\draw[-,thick,blue] (2.858 +0.625/2, -3.3/2 +0.625*0.866) to (2.858-0.625/2, -3.3/2 -0.625*0.866);
\draw[fill=gray, draw=none, fill opacity = 0.5] (2.858 +0.625/2, -3.3/2 +0.625*0.866) arc (60:-120:0.625) to (2.858-0.625/2, -3.3/2 -0.625*0.866);
\draw[-,thick,blue] (-2.858 -0.625/2, -3.3/2 +0.625*0.866) to (-2.858 +0.625/2, -3.3/2 -0.625*0.866);
\draw[fill=gray, draw=none, fill opacity = 0.5] (-2.858 -0.625/2, -3.3/2 +0.625*0.866) arc (-60-180:120-180:0.625) to (-2.858 +0.625/2, -3.3/2 -0.625*0.866);
\draw[-,thick,blue] (-2.858 -0.625/2, 3.3/2 -0.625*0.866) to (-2.858+0.625/2, 3.3/2 +0.625*0.866);
\draw[fill=gray, draw=none, fill opacity = 0.5] (-2.858 -0.625/2, +3.3/2 -0.625*0.866) arc (60-180:-120-180:0.625) to (-2.858+0.625/2, 3.3/2 +0.625*0.866);
%%horizons
\node at (0,3.3) {\textcolor{black}{$\cross$}};
\node at (2.857,1.65) {\textcolor{black}{$\cross$}};
\node at (2.857,-1.65) {\textcolor{black}{$\cross$}};
\node at (0,-3.3) {\textcolor{black}{$\cross$}};
\node at (-2.857,-1.65) {\textcolor{black}{$\cross$}};
\node at (-2.857,1.65) {\textcolor{black}{$\cross$}};
\node at (0,0) {\textcolor{black}{$\cross$}};
%%i's
\node at (0,2.5) {\textcolor{red}{$i_{1}$}};
\node at (2.857-0.8*0.866,1.65-0.8/2) {\textcolor{red}{$i_{2}$}};
\node at (2.857-0.8*0.866,-1.65+0.8/2) {\textcolor{red}{$i_{3}$}};
\node at (0,-3.3+0.8) {\textcolor{red}{$i_{4}$}};
\node at (-2.857+0.8*0.866,-1.65+0.8/2) {\textcolor{red}{$i_{5}$}};
\node at (-2.857+0.8*0.866,1.65-0.8/2) {\textcolor{red}{$i_{6}$}};
\end{tikzpicture}
\caption{\small An illustration of the connected configuration that contributes to $G^6_{i_{1}i_{2}i_{3}i_{4}i_{5}i_{6}}$.}
\label{pic:demo6}
\end{centering}
\end{figure}
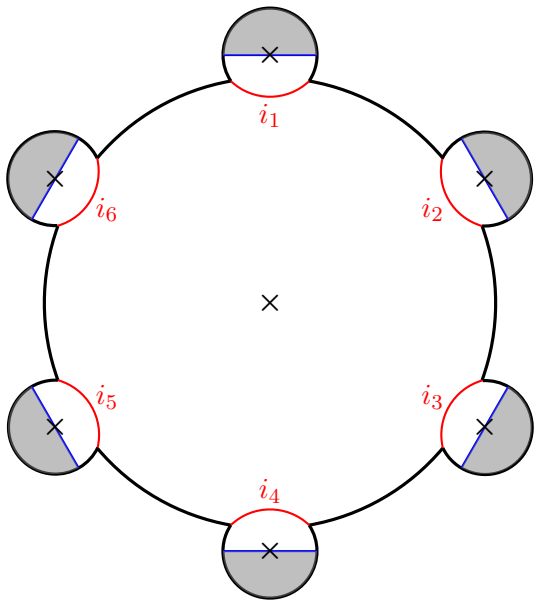

Let's firstly consider the higher moments of the overlap matrix
\begin{equation}
G^{n}_{i_{1}i_{2}\cdots i_{n}}\equiv G_{i_{1}i_{2}}G_{i_{2}i_{3}}\cdots G_{i_{n}i_{1}}\,,
\end{equation}
where $i_{1},i_{2},\cdots ,i_{n}\in \{1,2,\cdots\Omega$\}, for each $n\in\mathbb{Z}_{+}$. The higher moments $G^{n}_{i_{1}i_{2}\cdots i_{n}}$ can be computed using the saddle-point approximation of the gravitational path integral. We will firstly focus on the contribution from the fully connected saddle which we denote by
\begin{equation}
    G^{n}_{i_{1}i_{2}\cdots i_{n},c}\,.
\end{equation}
The bulk of the fully connected saddle has only one component whose interior connects all asymptotic boundaries (an example is Fig.~\ref{pic:demo6} for the case of $n=6$).\footnote{We don't consider topology changing processed in this paper as they should be suppressed for black holes with high temperature \cite{Penington:2019kki,Balasubramanian:2022gmo}.} The results of the disconnected saddles can be easily written down from connected saddles of lower moments with appropriate Kronecker-delta functions decorated. For example
\begin{equation}
G^{6}_{i_{1}i_{2}i_{3}i_{4}i_{5}i_{6}}\supset\delta_{i_{1}i_{4}}G^{3}_{i_{1}i_{2}i_{3}i_{1},c}G^{3}_{i_{1}i_{5}i_{6}i_{1},c}\,.
\end{equation}

The dust shell construction of the states $\ket{i}$ in the previous subsection enables us to compute these moments. These moments get contributions from connected configurations with matter shells connecting identical states (see Fig.~\ref{pic:demo6} for the configuration that contributes to $G^{6}_{i_{1}i_{2}i_{3}i_{4}i_{5}i_{6}}$). As we have discussed in Sec.~\ref{sec:approxstates}, in our case $m_{i}\rightarrow\infty$ and so the shells' contribution to the on-shell action is independent of the background geometry (see Appendix.~\ref{sec:shell} for details). Hence the contributions from the shells to $G^{n}_{i_{1}i_{2}\cdots i_{n},c}$ cancel between the numerator and the denominator. Moreover, the regions between the shells and the branes are suppressed so they don't contribute to the  $G^{n}_{i_{1}i_{2}\cdots i_{n},c}$ either. Therefore we have
\begin{equation}
    G^{n}_{i_{1}i_{2}\cdots i_{n},c}=\frac{Z(n\beta)}{Z(\beta)^{n}}\,.\label{eq:Gn1}
\end{equation}
So far we are focusing on black holes with fixed temperature $\beta$, nevertheless as we discussed in Sec.~\ref{sec:definition}, the black hole microstates we are interested in are in fact the microstates with energy close to the black hole mass $M$ in a small window with width $2\Delta E$, i.e.
\begin{equation}
    \bra{i}\hat{H}\ket{i}=E_{i}\in [M-\Delta E,M+\Delta E]\,,\quad \text{with }\Delta E\ll M\,.
\end{equation}
Thus we have
\begin{equation}
    Z(n\beta)=\int_{M-\Delta E}^{M+\Delta E} \rho(E)e^{-n\beta E}dE\approx 2e^{-n\beta M}\rho(M)\Delta E\,,
\end{equation}
and 
\begin{equation}
    Z(\beta)=\int_{M-\Delta E}^{M+\Delta E}\rho(E) e^{-n\beta E}dE\approx 2e^{-\beta M} \rho(M)\Delta E\,,
\end{equation}
where $2\rho(M)\Delta E$ can be computed using black hole thermodynamics as\footnote{That is we are using the first law of black hole thermodynamics \cite{Bardeen:1973gs}
\begin{equation}
    \log Z(\beta)=-\beta F=S-\beta E\,,
\end{equation}
with $E$ as the black hole mass and $S$ the Bekenstein-Hawking entropy Equ.~(\ref{eq:BH}).}
\begin{equation}
    2\rho(M)\Delta E=e^{S_{\text{BH}}}\,.
\end{equation}
As a result, we have
\begin{equation}
    G^{n}_{i_{1}i_{2}\cdots i_{n},c}=\frac{Z(n\beta)}{Z(\beta)^{n}}=e^{-(n-1)S_{\text{BH}}}\,.\label{eq:Gn2}
\end{equation}

\subsection{Extracting the Microscopic Entropy of the Black Hole}\label{sec:entropy}
In this subsection, we discuss how one can extract the microscopic entropy of the black hole microstates, i.e.~the dimensions of the Hilbert space of the black hole microstates using the higher moments $G^{n}_{i_{1}i_{2}\cdots i_{n}}$ we discussed in Sec.~\ref{sec:overlap}. 

In Sec.~\ref{sec:overlap}, we have taken $\Omega$ black hole microstates $\ket{i}$ from a continuous set of microstates. To prove the Bekenstein-Hawking entropy formula Equ.~(\ref{eq:BH}) we have to show that when $\Omega>\Omega_{\text{BH}}$ we only have $\Omega_{\text{BH}}$ linearly independent states among them. In other words, there should be $\Omega-\Omega_{\text{BH}}$ null states among the $\Omega$ microstates if $\Omega>\Omega_{\text{BH}}$. This can be proved borrowing the ideas from the random matrix theory \cite{Eynard:2015aea, Penington:2019kki, Balasubramanian:2022gmo} as follows. We first compute the resolvent matrix $W_{ij}$ of the overlapping matrix $G_{ij}$,
\begin{equation}
    W_{ij}(x)=\left(\frac{1}{x-G}\right)_{ij}=\frac{1}{x} \delta_{ij}+\sum_{n=1}^{\infty} \frac{1}{x^{n+1}}(G^n)_{ij}\,,
\end{equation}
where we defined $(G^{n})_{ij}=\sum_{i_{2},\cdots,i_{n}=1}^{\Omega}G_{ii_{2}}G_{i_{2}i_{3}}\cdots G_{i_{n}j}$. Then we can read off the density of eigenvalues for the overlap matrix $G_{ij}$ from the discontinuity of the resolvent,
\begin{equation}
    \rho(x)=\frac{1}{2\pi i}\Big(W(x-i0)-W(x+i0)\Big)\,,\label{eq:trick}
\end{equation}
where $W(x)=\Tr W_{ij}(x)=\sum_{i=1}^{\Omega}W_{ii}(x)$.

\begin{figure}
	\centering
	\includegraphics[width=1\linewidth]{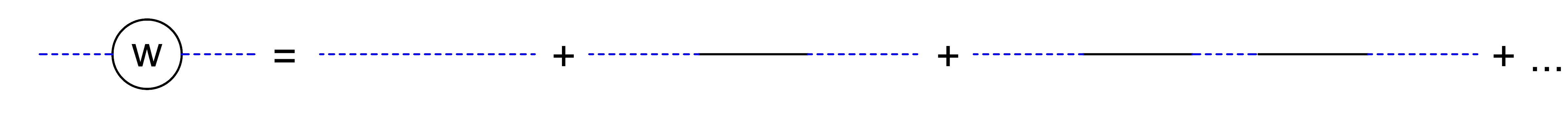} 
	 \caption{\small Diagrammatic representation of the resolvent matrix $W_{ij}$ on the CFT side.}
	\label{pic:sdequation1}
\end{figure}

\begin{figure}
	\centering
	\includegraphics[width=1\linewidth]{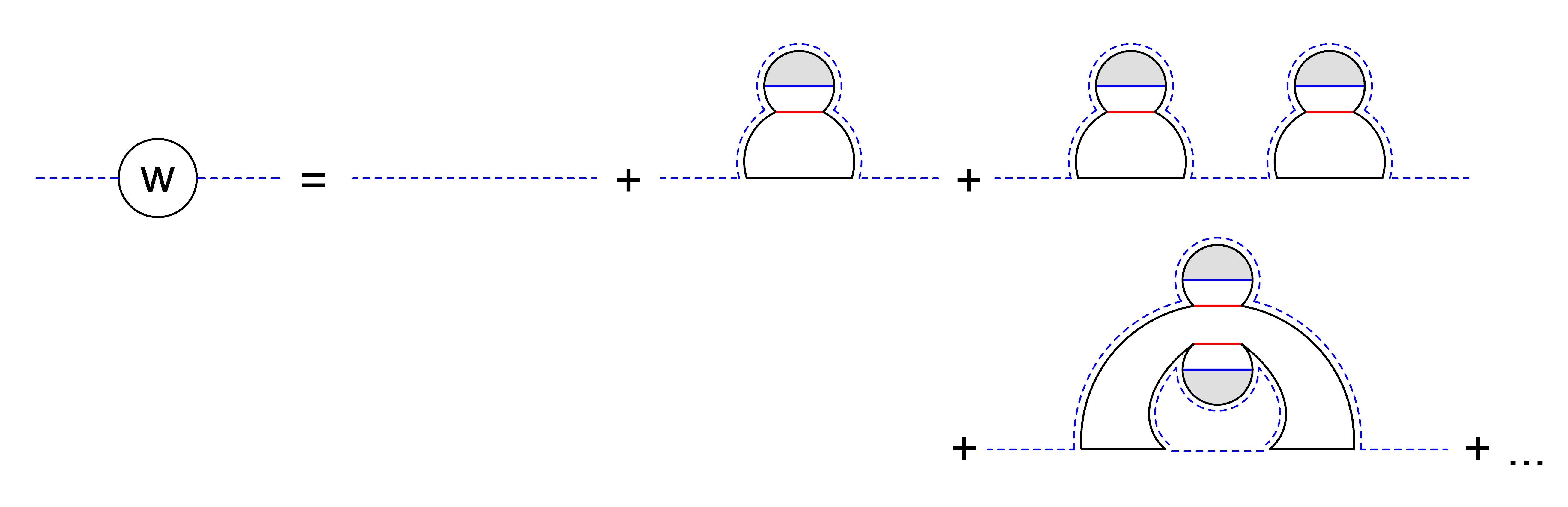}
  \caption{\small Diagrammatic representation of the resolvent matrix $W_{ij}$ on the gravity side.}
	\label{pic:sdequation2}
\end{figure}

\begin{figure}
	\centering
	\includegraphics[width=1\linewidth]{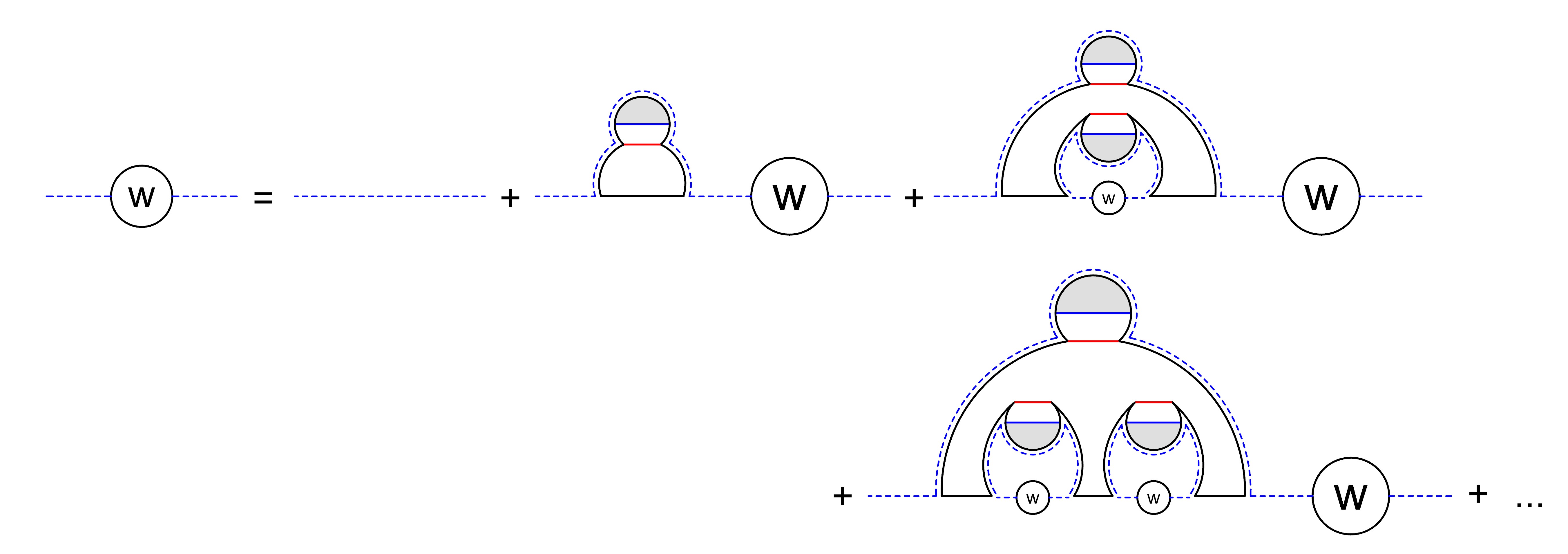}
  \caption{\small A reorganization of the expansion in Fig.~\ref{pic:sdequation2} which leads to the Schwinger-Dyson equation Equ.~(\ref{eq:sdequation}). }
	\label{pic:sdequation3}
\end{figure}

A diagrammatic representation of the matrix elements $W_{ij}$ in the CFT description is depicted in Fig.~\ref{pic:sdequation1}, where each dashed line comes with a factor of $1/x$, and the solid lines correspond to asymptotic AdS boundaries. In the gravitational description, we can compute the matrix elements $W_{ij}$ by filling the boundaries with all possible gravitational saddles in the bulk as in Fig.~\ref{pic:sdequation2} where we don't consider any topology change. The expansion depicted in Fig.~\ref{pic:sdequation2} can be exactly resumed by deriving a Schwinger-Dyson equation and solve it. We can reorganize the expansion depicted in Fig.~\ref{pic:sdequation2} into that depcited in Fig.~\ref{pic:sdequation3} where the sum is now ordered by the number of connected boundaries with all propagators as the resumed $W_{ij}$. More explicitly, we have
\begin{equation}
    \begin{split}
        W_{ij}(x)&=\frac{1}{x} \delta_{ij}+\frac{1}{x}\sum_{n=1}^{\infty} \Big(\sum_{i_{2},\cdots,i_{n}=1}^{\Omega}G^{n}_{ii_{2}\cdots i_{n},c} W_{i_{2}i_{2}}(x)\cdots W_{i_{n}i_{n}}\Big)W_{ij}(x)\,,\\&=\frac{1}{x}\delta_{ij}+\frac{1}{x}\sum_{n=1}^{\infty}\frac{Z(n\beta)}{Z(\beta)^{n}}W(x)^{n-1}W_{ij}(x)\,,\label{eq:sdequation}
    \end{split}
\end{equation} 
where we used Equ.~(\ref{eq:Gn1}). Taking the trace on both sides of the equation \eqref{eq:sdequation}, we get,
\be
\begin{aligned}
x W(x)&=\Omega +\sum_{n=1}^{\infty}\frac{Z(n\beta)}{Z(\beta)^n} W(x)^{n}\\
&=\Omega +\sum_{n=1}^{\infty}e^{-(n-1)S_{\text{BH}}} W(x)^{n}\\
&=\Omega+\frac{e^{S_{\text{BH}}} W(x)}{e^{S_{\text{BH}}}-W(x)}\,,\label{eq:SchwingerDyson}
\end{aligned}
\ee
where we used \eqref{eq:Gn2} and we resumed a geometrical series. This is the Schwinger-Dyson equation and we will solve for $W(x)$. The Schwinger-Dyson equation Equ.~(\ref{eq:SchwingerDyson}) leads to the following quadratic equation for $W(x)$,
\begin{equation}
    W(x)^2+\left(\frac{e^{S_{\text{BH}}}-\Omega}{x}-e^{S_{\text{BH}}}\right) W(x)+\frac{\Omega}{x} e^{S_{\text{BH}}}=0\,,\label{eq:quadratic}
\end{equation} 
which has the solutions
\begin{equation}
    W(x)=\frac{e^{S_{\text{BH}}}}{2}+\frac{\Omega-e^{S_{\text{BH}}}}{2x}\pm\frac{e^{S_{\text{BH}}}}{2x}\sqrt{\Big[x-(1-\Omega^{\frac{1}{2}}e^{-\frac{S_{\text{BH}}}{2}})^2\Big]\Big[x-(1+\Omega^{\frac{1}{2}}e^{-\frac{S_{\text{BH}}}{2}})^2\Big]}\,.\label{eq:solutionW}
\end{equation}
Now using Equ.~(\ref{eq:trick}), we get
\begin{equation}
\begin{split}
    \rho(x)=&\theta\Big((\sqrt{\Omega}e^{\frac{S_{\text{BH}}}{2}}-1)^{2},(\sqrt{\Omega}e^{\frac{S_{\text{BH}}}{2}}+1)^{2}\Big)(x)\frac{e^{S_{\text{BH}}}}{2\pi x}\sqrt{\Big[x-(\Omega^{\frac{1}{2}}e^{-\frac{S_{\text{BH}}}{2}}-1)^2\Big]\Big[(\Omega^{\frac{1}{2}}e^{-\frac{S_{\text{BH}}}{2}}+1)^{2}-x\Big]}\\&+\delta(x)(\Omega-e^{S_{\text{BH}}})\theta(\Omega-e^{S_{\text{BH}}})\,,\label{eq:rho}
    \end{split}
\end{equation}
where we choose the proper solution from Equ.~(\ref{eq:solutionW}) such that $\rho(x)>0$ for $x>0$. $\theta(a,b)(x)=1$ if $a<x<b$ and zero otherwise. As a consistency check, we have
\begin{equation}
    \int_{-\infty}^{\infty} dx\rho(x)=\Omega\,.\label{eq:checkconsistent}
\end{equation}
 We note that the density of eigenvalues $\rho(x)$ has a singular piece $\delta(x) (\Omega-e^{S_{\text{BH}}})\theta(\Omega-e^{S_{\text{BH}}})$, which tells us that the number of null states is $\Omega-e^{S_{\text{BH}}}$ if $\Omega>e^{S_{\text{BH}}}$. As a result, we can see that the number of linearly independent black hole microstates is
\begin{equation}
    \Omega_{\text{BH}}=\Omega-(\Omega-e^{S_{\text{BH}}})=e^{S_{\text{BH}}}\,,
\end{equation}
which is exactly the Bekenstein-Hawking formula Equ.~(\ref{eq:BH}). This finished our state-counting proof of the Bekenstein-Hawking entropy formula for single-sided black holes. Moreover, we also notice that the eigenvalue density $\rho(x)$ in Equ.~(\ref{eq:rho}) has zero support on $x<0$ which is consistent with unitarity. 


\section{Conclusions and Discussions}\label{sec:conclusion}
In this paper, we provided a state-counting derivation of the Bekenstein-Hawking entropy formula Equ.~(\ref{eq:BH}) for single-sided black holes in anti-de Sitter space. We focused on the case of $(2+1)$-dimensional BTZ black holes and our calculation is fully analytic. Our calculation is enabled by the black hole microstates we constructed in Sec.~\ref{sec:approxstates}. This set of states are legitimate black hole microstates following the definition in Sec.~\ref{sec:definition} and their construction allows us to extract the dimension of the black hole microstate Hilbert space by computing the moments of their overlap matrix $G_{ij}$. Our calculation and construction can be easily generalized to higher spacetime dimensions which though requires some numerics (see \cite{Cooper:2018cmb} for some relevant results). We want to emphasize that our calculation is done in the context of quantum general relativity, though the AdS/CFT correspondence is the guiding principle for our constructions,\footnote{Therefore, as opposed to constructions in \cite{Balasubramanian:2022lnw}, our construction of singled-sided black hole microstates has a clear CFT interpretation.} which doesn't require any specific constructions of black holes from strings or branes. Given the universality of the Bekenstein-Hawking entropy formula, we believe that a general microscopic understanding of this formula shouldn't rely on specific constructions of the black holes in an underlying UV complete theory. Therefore our construction and calculation captured these universal property of the Bekenstein-Hawking entropy. The basic lesson we learned from here is that the seemingly orthogonal large number of states, in fact, non-vanishingly overlap with each other, where the overlap is a nonperturbative gravitational effect and is enough to reproduce the Bekenstein-Hawking entropy formula.\footnote{It would be interesting to explore the distinguishability of the microstates we constructed by generalizing the studies in \cite{Bao:2017guc,Bao:2021ebo}. We thank Ning Bao for raising this point to us.} These are also the essence to avoid the bag-of-gold paradox. It is also interesting to note that our result Equ.~(\ref{eq:rho}) shows that there are no negative norm states which ensures the unitarity. Furthermore, going one-step further, if one wants to study the physics in the black hole interior especially close to the singularity, it would be helpful if there exists a precise construction of a black hole microstate where detailed calculations can be done using string theory. As far as we know, such a construction doesn't exist yet and the microscopic solutions constructed in the fuzzball program \cite{Giusto:2004id,Giusto:2004ip,Mathur:2005zp,Bena:2016ypk,Ganchev:2023sth} so far face the issue we discussed in Sec.~\ref{sec:definition} as they don't satisfy the definition Equ.~(\ref{eq:microstatedef}) of a black hole microstate \cite{Raju:2018xue}. Hence if one wants to study the detailed dynamics of black holes starting from these microscopic solutions, one has to take an average over a large number, at least $e^{S_{\text{BH}}}$, of them. This is a huge amount of power cost as one firstly has to do the calculation case by case and then take the average. Another challenge for this proposal is that so far there are not enough microscopic solutions constructed in the fuzzball program that one can average over and also it is not quite clear how the average should be carried out. We believe that these are interesting questions that have to be studied in the future for a better understanding of black holes.

\section*{Acknowledgements}
We are grateful to Ning Bao, Roberto Emparan, Stefano Giusto, Andreas Karch, Joe Minahan, Suvrat Raju, Lisa Randall and Martin Sasieta for relevant discussions. We thank Roberto Emparan for helpful comments on the draft. We thank Zhenhao Zhou for plotting some of the diagrams in this paper. HG would like to thank the hospitality from the Aspen Center for Physics where the final stage of this work is performed. Research at the Aspen Center for Physics is supported by National Science Foundation grant PHY-2210452 and a grant from the Simons Foundation (1161654, Troyer). The work of HG is supported by a
grant from Physics Department at Harvard University. The work of YJ is supported by the U.S Department of Energy ASCR EXPRESS grant, Novel Quantum Algorithms from Fast Classical Transforms, and Northeastern University.

\appendix
\section{Detailed Analysis of the Brane}\label{sec:brane}

In this section, we perform a detailed analysis of the configuration of the Karch-Randall brane that is behind the black hole horizon.\footnote{See \cite{Cooper:2018cmb,Antonini:2019qkt,Antonini:2021xar,Antonini:2022blk,Waddell:2022fbn,Antonini:2023hdh,Antonini:2024bbm} for some interesting explorations of this model in various contexts.} This is the basic set-up that we considered in the main text. We also calculate the on-shell action of the brane configurations for $d=2$ which is frequently used in the main text.

\subsection{The Configuration of the Brane}
In the Euclidean signature, the brane lives in the black hole geometry 
\begin{equation}
    ds^2=f(r)d\tau^2+\frac{dr^2}{f(r)}+r^2 d\Omega_{d-1}^2\,,\quad f(r)=1+r^{2}-\frac{16\pi G_{N}M}{(d-1)\Omega_{d-1}r^{d-2}}\,,\label{eq:BHmetricE}
\end{equation}
where $G_{N}$ is Newton's constant, $M$ is the mass parameter of the black hole and we have set the AdS length scale $l_{\text{AdS}}$ to be one. The brane configuration is determined by the brane embdedding equation
\begin{equation}
    h_{ab}K-K_{ab}= T h_{ab}\,,\label{eq:braneeom}
\end{equation}
where $T$ is the tension of the brane, $h_{ab}$ is the induced metric on the brane, $K_{ab}$ is the extrinsic curvature of the brane and $K$ is its trace. The brane is a codimension one object which obeys the bulk spherical symmetry, i.e.~it wraps the $S^{d-1}$. Hence we only have to consider the $\tau$ and $r$ directions. Let's take the proper length in these two directions on the brane to be $l$, i.e.~we have
\begin{equation}
    f(r)\dot{\tau}^{2}(l)+\frac{\dot{r}^{2}(l)}{f(r)}=1\,.\label{eq:tangent}
\end{equation}
The extrinsic curvature of the brane can be computed from its inward pointing unit normal vector. The inward pointing unit normal vector on the brane is given by
\begin{equation}
    n_{a}=(-\dot{r}(l),\dot{\tau}(l),\vec{0})\,,
\end{equation}
where the first component is the $\tau$-component, the second component is the $r$-component and the normal vector has zero components along the $S^{d-1}$. The extrinsic curvature can be calculated using 
\begin{equation}
    K_{ab}=-h_{a}^{c}h_{b}^{d}\nabla_{c}n_{d}\,.
\end{equation}
As a result, we have
\begin{equation}
    K_{ll}=\frac{\frac{d}{dl}\sqrt{f(r)-\dot{r}^{2}(l)}}{\dot{r}(l)}\,,\quad K_{\Omega_{i}\Omega_{j}}=r(l)\sqrt{f(r)-\dot{r}^{2}(l)}\omega_{\Omega_{i}\Omega_{j}}\,,
\end{equation}
where for the first equation one has to use $K_{ll}=t^{\mu}t^{\nu}\nabla_{\mu}n_{\nu}$ with $t^{\mu}$ is the unit tangent vector and $\omega_{\Omega_{i}\Omega_{j}}$ denotes the metric for a $S^{d-1}$ with unit radius. Thus the brane embedding equation Equ.~(\ref{eq:braneeom}) can be reduced to
\begin{equation}
    \begin{split}
        \Big[\frac{\frac{d}{dl}\sqrt{f(r)-\dot{r}^{2}(l)}}{\dot{r}(l)}+\frac{d-1}{r(l)}\sqrt{f(r)-\dot{r}^{2}(l)}\Big]-\frac{1}{r(l)}\sqrt{f(r)-\dot{r}^{2}(l)}&=T\,,\\\Big[\frac{\frac{d}{dl}\sqrt{f(r)-\dot{r}^{2}(l)}}{\dot{r}(l)}+\frac{d-1}{r(l)}\sqrt{f(r)-\dot{r}^{2}(l)}\Big]-\frac{\frac{d}{dl}\sqrt{f(r)-\dot{r}^{2}(l)}}{\dot{r}(l)}&=T\,,
    \end{split}
\end{equation}
which gives
\begin{equation}
    \dot{r}(l)^2=f(r)-\frac{T^2}{(d-1)^2}r^2\,,\label{eq:brane2}
\end{equation}
where the tension of the Karch-Randall brane is always smaller than the critical tension $T_{c}=d-1$. Using Equ.~(\ref{eq:tangent}) and Equ.~(\ref{eq:brane2}), we can compute the time the brane spends in the bulk
\begin{equation}
    \Delta\tau=2\int_{r_{c}}^{\infty}\frac{dr}{f(r)}\sqrt{\frac{\frac{T^2}{(d-1)^2}r^2}{f(r)-\frac{T^2}{(d-1)^2}r^2}}\,,\label{eq:taubrane}
\end{equation}
where $r_{c}$ is the critical point, i.e.~$f(r_c)-\frac{T^2}{(d-1)^{2}}r_{c}^{2}=0$. In general dimensions both $r_{c}$ and $\Delta\tau$ have to be found numerically. However when $d=2$ they can be computed analytically. In $d=2$ we have 
\begin{equation}
    r_{c}=r_{H}\frac{1}{\sqrt{1-T^2}}\,,\quad \Delta\tau=\frac{\beta}{2}\,,\label{eq:braneconstraint}
\end{equation}
where we defined the black hole radius $r_{H}^2=8 G_{N}M-1$ in terms of which the inverse Hawking temperature is given by $\beta=\frac{2\pi}{r_{H}}$. In higher dimensions one can show that $\Delta\tau>\frac{\beta}{2}$.

\subsection{On-Shell Action and Its Renormalization}
In this subsection, we will evaluate the on-shell value of the action Equ.~(\ref{eq:action1}) in the Euclidean signature solutions Equ.~(\ref{eq:BHmetricE}) with the brane Equ.~(\ref{eq:brane2}). This quantity is used to determine various state-overlaps using the saddle-point approximation for the gravitational path integral.

The cosmological constant in $d=2$ with $l_{\text{AdS}}=1$ is given by $\Lambda=-1$ and the on-shell value of the Ricci scalar is $R=-6$. The brane trajectory can be solved as
\begin{equation}
    r(\tau)=\frac{1}{\sqrt{1-T^2}}\sqrt{T^{2}r_{H}^{2}\tan^{2}(r_{H}\tau)+r_{H}^{2}}\,,
\end{equation}
and trace of its extrinsic curvature is $K=2T$. The trace of the extrinsic curvature of asymptotic boundary is 
\begin{equation}
K=\sqrt{f(r_{\epsilon})}\Big(\frac{f'(r_{\epsilon})}{2f(r_{\epsilon})}+\frac{1}{r_{\epsilon}}\Big)\,,
\end{equation}
where we take the asymptotic boundary to be at the cutoff surface $r=r_{\epsilon}$ and we will take $r_{\epsilon}\rightarrow\infty$ at the end of the calculation. The on-shell action can be calculated by cutting the bulk into two pieces as in Fig.~\ref{pic:cut}, evaluating the action in the two pieces separately and summing them up at the end. The piece that contains the brane contributes to the on-shell action
\begin{equation}
\begin{split}
    S_{\text{green}}&=-\frac{1}{16\pi G_{N}}\int_{\text{green}}d^{3}x\sqrt{g}(R-2\Lambda)-\frac{1}{8\pi G_{N}}\int_{\mathcal{B}_{2}} d^{2}x\sqrt{h}(K-T)\,,\\&=-\frac{1}{16\pi G_{N}} (-4)(2\pi)\int_{-\frac{\beta}{4}}^{\frac{\beta}{4}}d\tau\int_{r_{H}}^{r(\tau)} rdr-\frac{T}{8\pi G_{N}}(2\pi)\int_{-\frac{\beta}{4}}^{\frac{\beta}{4}}\frac{r^{2}(\tau)-r_{H}^{2}}{T r(\tau)} d\tau r(\tau)\,,\\&=0\,.
    \end{split}
\end{equation}
The piece that contains the asymptotic boundary contributes to the on-shell action
\begin{equation}
\begin{split}
    S_{\text{orange}}&=-\frac{1}{16\pi G_{N}}\int_{\text{orange}}d^{3}x\sqrt{g}(R-2\Lambda)-\frac{1}{8\pi G_{N}}\int_{\partial\mathcal{M}_{2}} d^{2}x\sqrt{h}K\,,\\&=-\frac{1}{16\pi G_{N}}(-4)(2\pi)\frac{\beta}{2}\int_{r_{H}}^{r_{\epsilon}}rdr-\frac{1}{8\pi G_{N}}(2\pi)\frac{\beta}{2}r_{\epsilon}f(r_{\epsilon})\Big(\frac{f'(r_{\epsilon})}{2f(r_{\epsilon})}+\frac{1}{r_{\epsilon}}\Big)\,,\\&=\frac{4\pi\beta}{16\pi G_{N}}\frac{r_{\epsilon}^{2}-r_{H}^{2}}{2}-\frac{\beta}{8G_{N}}(2r_{\epsilon}^{2}-r_{H}^{2})+\mathcal{O}(\frac{1}{r_{\epsilon}^{2}})\,.
    \end{split}
\end{equation}
As a result, we have the on-shell action
\begin{equation}
    S_{\text{on-shell}}=S_{\text{green}}+S_{\text{orange}}=-\frac{\beta  r_{\epsilon}^{2}}{8G_{N}}+\mathcal{O}(\frac{1}{r_{\epsilon}^{2}})\,.\label{eq:os}
\end{equation}
The renormalized on-shell action is given by $S_{\text{on-shell}}+S_{\text{ct}}$ where the counter term in AdS$_{3}$ is given by \cite{Balasubramanian:1999re}\footnote{We note that the definition of the extrinsic curvature in \cite{Balasubramanian:1999re} is opposite to ours so the counter term is also opposite.}
\begin{equation}
    S_{\text{ct}}=\frac{1}{8\pi G_{N}}\int_{\partial\mathcal{M}_{2}}d^{2}x\sqrt{h}=\frac{1}{8\pi G_{N}}(2\pi)\frac{\beta}{2}r_{\epsilon}\sqrt{f(r_{\epsilon})}=\beta\frac{r_{\epsilon}^{2}-\frac{1}{2}r_{H}^{2}}{8 G_{N}}+\mathcal{O}(\frac{1}{r_{\epsilon}^{2}})\,.\label{eq:ct}
\end{equation}
Finally, adding up the on-shell action Equ.~(\ref{eq:os}) and the counter term Equ.~(\ref{eq:ct}) and taking the $r_{\epsilon}\rightarrow\infty$ limit, we have the renormalized on-shell action
\begin{equation}
    S^{\text{ren}}=-\frac{\beta}{16 G_{N}}r_{H}^{2}=-\frac{\pi^{2}}{4\beta G_{N}}\,,
\end{equation}
where we used $\beta=\frac{2\pi}{r_{H}}$. We note that $S^{\text{ren}}$ is independent of the brane tension. We will denote $e^{-S^{\text{ren}}}$ as $\sqrt{Z(\beta)}$ and the normalization of the states $\ket{\Psi_{i}(\beta)}$ in the main text is
\begin{equation}
\bra{\Psi_{i}(\beta)}\ket{\Psi_{i}(\beta)}=\sqrt{Z(\beta)}\,,
\end{equation}
for $\forall i$.

\begin{figure}
\begin{centering}
    \begin{tikzpicture}[scale=1]
    \draw[-,very thick,black] (2.5,0) arc (0:360:2.5);
    \draw[-,very thick,blue] (0,2.5) arc (150:210:5);
     \draw[fill=gray, draw=none, fill opacity = 0.5] (0,2.5) arc (150:210:5) arc (-90:-270:2.5);
\node at (0,0) {\textcolor{black}{$\cross$}};
  \draw[fill=green, draw=none, fill opacity = 0.5] (0,2.5) arc (150:210:5) to (0,2.5);
   \draw[fill=orange, draw=none, fill opacity = 0.5] (0,2.5) to (0,-2.5) arc (-90:90:2.5);
     \end{tikzpicture}
\caption{}
\label{pic:cut}
\end{centering}
\end{figure}

\subsection{Comparing with the CFT Partition Function}
As a nontrivial check of the duality between the bulk description of the state $\ket{\Psi_{i}(\beta)}$ and the CFT description, let's compute
\begin{equation}
    \bra{\Psi_{i}(\beta)}\ket{\Psi_{i}(\beta)}=\bra{B}e^{-\frac{\beta}{2}\hat{H}}\ket{B}\,,
\end{equation}
in the high temperature limit using the CFT$_{2}$ description. This can be computed using the modular invariance or open-closed duality in CFT$_{2}$ \cite{Cardy:2004hm, Polchinski:1998rq} which gives
\begin{equation}
\begin{split}
    \bra{\Psi_{i}(\beta)}\ket{\Psi_{i}(\beta)}=\bra{B}e^{-\frac{\beta}{2}\hat{H}}\ket{B}=\Tr_{B}e^{-\frac{4\pi}{\beta}\hat{H}_{\text{open}}}\,,
    \end{split}
\end{equation}
where in the last step we are taking the trace of the states of a CFT$_{2}$ living on an interval with length $1$\footnote{That is the modulus of the cylinder $t=\frac{2\pi}{(\frac{\beta}{2})}=\frac{4\pi}{\beta}$ is fixed.} and two boundaries specified by the boundary condition $B$ and $\hat{H}_{\text{open}}=\pi\hat{L}_{0}-\frac{\pi c}{24}$ with $\hat{L}_{0}$ as the zeroth Virasoro symmetry generator. In the high temperature limit we have
\begin{equation}
     \bra{\Psi_{i}(\beta)}\ket{\Psi_{i}(\beta)}=\Tr_{B}e^{-\frac{4\pi}{\beta}\hat{H}_{\text{open}}}=e^{\frac{\pi^{2} c}{6\beta}} \,,
\end{equation}
which is exactly the same as $e^{-S^{\text{ren}}}$ if we use the Brown-Henneaux central charge $c=\frac{3}{2 G_{N}}$ \cite{Brown:1986nw}.

\section{Detailed Analysis of the Shell}\label{sec:shell}
In this section, we perform an analysis of the trajectory of the shell, its backreaction to the bulk geometry and the large mass limit of the shell. The goal is to derive useful formulas for the discussions in the main text. 
\subsection{The Configuration of the Shell}
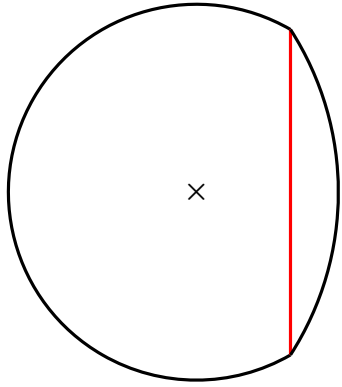
\begin{figure}
\begin{centering}
    \begin{tikzpicture}[scale=1]
    \draw[-,very thick,black] (1.25,2.165) arc (60:300:2.5);
    \draw[-,very thick,red] (1.25,2.165) to (1.25,-2.165);
\node at (0,0) {\textcolor{black}{$\cross$}};
  \draw[-,very thick,black] (1.25,2.165) arc (32.769:-32.769:4);
     \end{tikzpicture}
\caption{The red curve indicates the world volume of the shell and the cross is the horizon of the geometry of the left black hole of the shell. The black hole geometry on the left of the shell is of mass $M_{-}$ and that on its right is of mass $M_{+}$.}
\label{pic:gluing1}
\end{centering}
\end{figure}

\begin{figure}
\begin{centering}
    \begin{tikzpicture}[scale=0.8]
    \draw[-,very thick,black] (1.25,2.165) arc (60:300:2.5);
    \draw[-,very thick,red] (1.25,2.165) to (1.25,-2.165);
\node at (0,0) {\textcolor{black}{$\cross$}};
  \draw[-,very thick,black] (1.25,2.165) arc (180-46.192:-180+46.192:3);
  \node at (1.25+2.077,0) {\textcolor{black}{$\cross$}};
     \end{tikzpicture}
\caption{The red curve indicates the world volume of the shell and the crosses are the horizon of the respective geometries of the black holes on the left and right side of the shell. The black hole geometry on the left of the shell is of mass $M_{-}$ and that on its right is of mass $M_{+}$.}
\label{pic:gluing2}
\end{centering}
\end{figure}
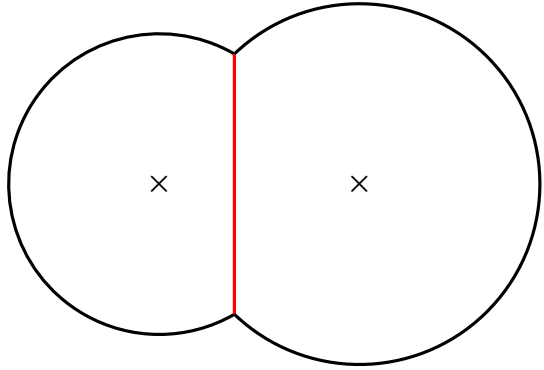

The shell glues the following two black hole geometries together
\begin{equation}
    ds^{2}_{\pm}=f_{\pm}(r_{\pm})d\tau^{2}_{\pm}+\frac{dr^{2}_{\pm}}{f_{\pm}(r_{\pm})}+r^{2}_{\pm}d\Omega^{2}_{d-1}\,,
\end{equation}
in a continuous manner. In the above geometry we have the blackening factor
\begin{equation}
    f_{\pm}(r)=1+r^2-\frac{16\pi G_{N}M_{\pm}}{(d-1)\Omega_{d-1}r^{d-2}}\,,\label{eq:blackening}
\end{equation}
where $G_{N}$ is Newton's constant and $M$ is the mass of the black hole. Let's take the proper length on the shell to be $s$ and consider a spherically symmetric shell, i.e.
\begin{equation}
    f_{\pm}(r_{\pm}(s))\dot{\tau}^{2}_{\pm}(s)+\frac{\dot{r}^{2}_{\pm}(s)}{f_{\pm}(r_{\pm}(s))}=1\,.\label{eq:normalization}
\end{equation}
The continuity of the ambient geometry in the spherical directions along the shell requires
\begin{equation}
    r_{+}(s)=r_{-}(s)\,.\label{eq:cont}
\end{equation}
Moreover, the vanishing of the variation of the action requires the Israel's junction condition across the shell
\begin{equation}
    h_{ab}\Delta K-\Delta K_{ab}=8\pi G_{N}T_{ab}\,,\label{eq:junction}
\end{equation}
where $\Delta K_{ab}=K_{ab,+}-K_{ab,-}$ is the jump of the extrinsic curvature of the shell, $\Delta K$ is its trace, $h_{ab}$ is the induced metric on the shell and $T_{ab}$ is the energy-momentum tensor of the shell. As we discussed below Equ.~(\ref{eq:shelleom}), the shell energy-momentum tensor is given by $T_{ab}=\sigma(s) u_{a}u_{b}$ where $\sigma(s)$ is the mass density of the shell and $u_{a}$ is its unitly normalized velocity vector. The extrinsic curvatures can be computed from the inward pointing unit normal vectors using the formula
\begin{equation}
    K_{ab,\pm}=-h_{a,\pm}^{c}h_{b,\pm}^{d} \nabla_{c}^{\pm}n^{\pm}_{d}\,,
\end{equation}
where $h_{ab}$ is the induced metric on the shell and the indices are lifted using the inverse bulk metric $g_{\pm}^{ab}$. There are two possible configurations as depicted in Fig.~\ref{pic:gluing1} and Fig.~\ref{pic:gluing2}. If we choose the $\tau=0$ to be on the far right point along the asymptotic boundary (the black curves of Fig.~\ref{pic:gluing1} and Fig.~\ref{pic:gluing2}), the sign of  each component of the inward pointing normal vector $n^{\pm}_{a}$ is independent of the specific configuration of the gluing. The unit normal vectors are given as
\begin{equation}
    n^{\pm}_{a}=(\mp\dot{r}_{\pm}(s),\pm\dot{\tau}_{\pm}(s),\vec{0})\,,
\end{equation}
where the first component is in the $\tau$-direction, the second component is in the $r$-direction and the normal vector is not along the spherical directions at all. 

As a result, we have the nonzero components of the extrinsic curvature
\begin{equation}
    K_{ss,\pm}=\mp\frac{\frac{d}{ds}\sqrt{f_{\pm}(r_{\pm}(s))-\dot{r}_{\pm}^2(s)}}{\dot{r}_{\pm}(s)}\,,\quad K_{\Omega_{i}\Omega_{j},\pm}=\mp r_{\pm}(s)\sqrt{f_{\pm}(r_{\pm}(s))-\dot{r}_{\pm}^2(s)}\omega_{\Omega_{i}\Omega_{j}}\,,
\end{equation}
where for the first equation one has to use $K_{ss}=u^{\mu}u^{\nu}\nabla_{\mu}n_{\nu}$ with $u^{\mu}$ is the unit tangent vector and $\omega_{\Omega_{i}\Omega_{j}}$ denotes the metric for a $S^{d-1}$ with unit radius. Thus the junction condition Equ.~(\ref{eq:junction}) is reduced to
\begin{equation}
    \begin{split}
\frac{(d-1)}{r(s)}\Big[\sqrt{f_{+}(r(s))-\dot{r}^{2}(s)}+\sqrt{f_{-}(r(s))-\dot{r}^{2}(s)}\Big]&=8\pi G_{N}\sigma\,,\\\frac{1}{\dot{r}(s)}\frac{d}{ds}\Big[\sqrt{f_{+}(r(s))-\dot{r}^{2}(s)}+\sqrt{f_{-}(r(s))-\dot{r}^{2}(s)}\Big]&=8\pi G_{N}\frac{2-d}{d-1}\sigma\,,
    \end{split}
\end{equation}
where we have used Equ.~(\ref{eq:cont}), i.e.~$r_{+}(s)=r_{-}(s)$. Taking the ratio of the above two equations and integrating the resulting equation we get
\begin{equation}
    \sqrt{f_{+}(r(s))-\dot{r}^{2}(s)}+\sqrt{f_{-}(r(s))-\dot{r}^{2}(s)}=\frac{8\pi G_{N}m}{(d-1)\Omega_{d-1}r^{d-2}}\,, \label{eq:fshift}
\end{equation}
where $m=\Omega_{d-1}r^{d-1}(s)\sigma(s)$ is the mass of the shell. We can solve Equ.~(\ref{eq:fshift}) for the shell trajectory
\begin{equation}
    \dot{r}^{2}(s)=-V_{\text{eff}}(r)\,,\label{eq:rV}
\end{equation}
where we have defined
\begin{equation}
    -V_{\text{eff}}(r)=1+r^2-\frac{16\pi G_{N}(M_{+}+M_{-})}{2(d-2)\Omega_{d-1}r^{d-2}}-\Big(\frac{4\pi G_{N}m}{(d-1)\Omega_{d-1}r^{d-2}}\Big)^2-\Big(\frac{M_{+}-M_{-}}{m}\Big)^2\,.
\end{equation}
Using Equ.~(\ref{eq:normalization}) and Equ.~(\ref{eq:rV}), we can compute how much time the shell travels in the bulk and it is given by
\begin{equation}
\Delta\tau_{\pm}=2\int_{r_{c}}^{\infty}\frac{dr}{f_{\pm}(r)}\sqrt{\frac{V_{\text{eff}}(r)+f_{\pm}(r)}{-V_{\text{eff}}(r)}}\,,
\end{equation}
where $r_{c}$ is the critical point, i.e.~$V(r_{c})=0$. In general dimensions both $r_{c}$ and $\Delta\tau_{\pm}$ have to be found numerically. However when $d=2$ they can be computed analytically and one can check that the qualititive dependence of $\Delta\tau_{\pm}$ and $r_{c}$ on $M_{\pm}$ and $m$ is independent of the dimension $d$. In $d=2$ we have
\begin{equation}
    \begin{split}
        r_{c}=\sqrt{-1+8 G_{N}M_{-}+(2G_{N}m+\frac{M_{+}-M_{-}}{m})^2}\,,\quad\Delta\tau_{\pm}=\frac{\beta_{\pm}}{\pi}\arcsin(\frac{r_{H\pm}}{r_{c}})\,,\label{eq:shellconstraint}
    \end{split}
\end{equation}
where we have defined the black holes radius $r_{H\pm}^{2}=8 G_{N}M_{\pm}-1$ and the black hole inverse temperature is given by $\beta_{\pm}=\frac{2\pi}{r_{\pm}}$. In our paper, we only consider the situations that $M_{+}\geq M_{-}$.

\subsection{On-shell Action and Its Renormalization}
In this subsection, we will compute the contribution from the shell to the on-shell action in the $d=2$ case. In this case, we have
\begin{equation}
    \frac{dr_{\pm}}{d\tau_{\pm}}=(r_{\pm}^{2}-r_{H\pm}^{2})\sqrt{\frac{r_{\pm}^{2}-r_{c}^{2}}{r_{c}^{2}-r_{H\pm}^{2}}}\,,
\end{equation}
solving which we get
\begin{equation}
    r_{\pm}(\tau_{\pm})=\frac{r_{\pm}r_{c}\cos(r_{\pm}\tau_{\pm})}{\sqrt{r_{\pm}^{2}-r_{c}^{2}\sin^{2}(r_{\pm}\tau_{\pm})}}\,.
\end{equation}
However, to compute the shell contribution to the on-shell action it is easiest to use the proper length $s$ and then reparametrize it by $r_{\pm}$ and Equ.~(\ref{eq:rV})
\begin{equation}
    \begin{split}
S_{\text{shell}}&=\int_{\mathcal{S}}d^{2}x\sqrt{h}\sigma=2\pi\int ds  r(s)\sigma(s),\\&=2\pi \int\frac{1}{\dot{r_{\pm}}(s)}r_{\pm}\sigma dr_{\pm} =2\int_{r_{c}}^{r_{\epsilon}} dr_{\pm} \frac{1}{\sqrt{r_{\pm}^{2}-r_{c}^{2}}}(2\pi r_{\pm})\sigma\,\\&=2m\int_{r_{c}}^{r_{\epsilon}} dr_{\pm} \frac{1}{\sqrt{r_{\pm}^{2}-r_{c}^{2}}}=2m\log\frac{r_{\epsilon}+\sqrt{r_{\epsilon}^{2}-r_{c}^{2}}}{r_{c}}\,,\\&=2m\log\frac{2r_{\epsilon}}{r_{c}}\,,\label{eq:Sshell}
    \end{split}
\end{equation}
where in the last step we used the fact that $r_{\epsilon}\rightarrow\infty$. We can renormalize Equ.~(\ref{eq:Sshell}) by adding the counter term $-2m\log(2r_{\epsilon})$ and the renormalized action becomes
\begin{equation}
S_{\text{shell}}^{\text{ren}}=-2m\log r_{c}\,.\label{eq:Sshellren}
\end{equation}
In the large mass limit $m\gg M_{\pm}$ we have
\begin{equation}
S_{\text{shell}}^{\text{ren}}\approx-2m\log 2G_{N}m\,,\label{eq:Sshellren}
\end{equation}
where we have used Equ.~(\ref{eq:shellconstraint}) and this result is independent of the background geometry \cite{Sasieta:2022ksu}.

\bibliographystyle{JHEP}
\bibliography{main}
\end{sloppypar}
\end{document}